# THE RETRIEVAL PHASE OF THE HOPFIELD MODEL:
# A RIGOROUS ANALYSIS OF THE OVERLAP DISTRIBUTION#

Anton Bovier [1]

*Weierstraß–Institut*
*für Angewandte Analysis und Stochastik*
*Mohrenstraße 39, D-10117 Berlin, Germany*

Véronique Gayrard[2]

*Centre de Physique Théorique - CNRS*
*Luminy, Case 907*
*F-13288 Marseille Cedex 9, France*

**Abstract:** Standard large deviation estimates or the use of the Hubbard-Stratonovich transformation reduce the analysis of the distribution of the overlap parameters essentially to that of an explicitly known random function $\Phi_{N,\beta}$ on $I\!\!R^M$. In this article we present a rather careful study of the structure of the minima of this random function related to the retrieval of the stored patterns. We denote by $m^*(\beta)$ the modulus of the spontaneous magnetization in the Curie-Weiss model and by $\alpha$ the ratio between the number of the stored patterns and the system size. We show that there exist strictly positive numbers $0 < \gamma_a < \gamma_c$ such that 1) If $\sqrt{\alpha} \leq \gamma_a(m^*(\beta))^2$, then the absolute minima of $\Phi$ are located within small balls around the points $\pm m^* e^\mu$, where $e^\mu$ denotes the $\mu$-th unit vector while 2) if $\sqrt{\alpha} \leq \gamma_c(m^*(\beta))^2$ at least a local minimum surrounded by extensive energy barriers exists near these points. The random location of these minima is given within precise bounds. These are used to prove sharp estimates on the support of the Gibbs measures.

---

\# Work partially supported by the Commission of the European Communities under contract CHRX-CT93-0411
[1] e-mail: bovier@iaas-berlin.d400.de
[2] e-mail: gayrard@cpt.univ-mrs.fr

# I. Introduction

Over the last few years the so-called Hopfield model of an autoassociative memory [Ho], originally introduced by Figotin and Pastur [FP] as a simplified model of a spin glass, has emerged as one of the more interesting models for spin systems with strongly disordered interactions. (for a survey mathematical results on this model and related topics, see the lecture notes of Petritis [P]). In a series of recent papers we have, partly in collaboration with Pierre Picco, obtained a fairly complete understanding of the thermodynamic properties of the Hopfield model in the regime there the ratio of the number of patterns $M(N)$ and the number of neurons, $N$, tends to zero [BGP1,BG2], and even if $\lim \frac{M}{N} = \alpha > 0$, for very small $\alpha$, we have been able to prove the existence of disjoint Gibbs states corresponding to the different patterns at sufficiently low temperatures [BGP2]. Technically, this relied on the analysis in some way or the other on large deviation estimates for the distribution of the overlap parameters.

The purpose of the present note is to present a more refined analysis of these large deviation estimates intended for a more detailed investigation of its critical points and its behaviour near them in the case where $\alpha$ is strictly positive, though small. These are relevant not only for the analysis of the Gibbs states (where only the absolute minima are important) but also for the characterization of the long-time characteristics of the stochastic retrieval dynamics of the system. From numerical experiments and the replica heuristic it is expected that local minima of the "free energy functional" persist for considerably larger values of $\alpha$ than those for which they are *absolute* minima [AGS]. The 'storage capacity' is usually defined as the maximal value of $\alpha$ for which the *local* minima near the patterns exist. Newman [N], in a seminal paper of 1988 has proven a lower bound for the critical $\alpha$ for zero temperature (see also [KPa]). One of the main results of the present paper is an extension of this finding to positive temperatures. In particular, we give estimates on the behaviour of the critical $\alpha$ as a function of the temperature that show the expected power law behaviour near $T = 1$. Furthermore, we will compute rather precisely the exact (random) location of these minima and we will show that, for $T$ not too small, the rate function near the location of the original patterns is locally convex, implying that there exists a *unique* local minimum near the patterns. Moreover, we will show that the only macroscopic component of the overlap vector at the minima is (at $T \approx 0$) shifted down from one by a term of order $\exp(-1/(2\alpha))$, as predicted in [AGS].

Let us recall the definitions of the Hopfield model and the main quantities of interest. Let $\mathcal{S}_N \equiv \{-1,1\}^N$ denote the set of functions $\sigma : \{1,\ldots,N\} \to \{-1,1\}$, and set $\mathcal{S} \equiv \{-1,1\}^{I\!\!N}$. We call $\sigma$ a spin configuration and denote by $\sigma_i$ the value of $\sigma$ at $i$. Let $(\Omega, \mathcal{F}, I\!\!P)$ be an abstract probability space and let $\xi_i^\mu$, $i, \mu \in I\!\!N$, denote a family of independent identically distributed random variables on this space. For the purposes of this paper we will assume that $I\!\!P[\xi_i^\mu = \pm 1] = \frac{1}{2}$, but more general distributions can be considered. We will write $\xi^\mu[\omega]$ for the $N$-dimensional random vector whose $i$-th component is given by $\xi_i^\mu[\omega]$ and call such a vector a 'pattern'. On the other hand, we use the notation $\xi_i[\omega]$ for the $M$-dimensional vector with the same components. When we write $\xi[\omega]$ without indices, we frequently will consider it as an $M \times N$ matrix and we write $\xi^t[\omega]$ for the transpose of this matrix. Thus, $\xi^t[\omega]\xi[\omega]$ is the $M \times M$ matrix whose elements are $\sum_{i=1}^N \xi_i^\mu[\omega]\xi_i^\nu[\omega]$. With this in mind we will use throughout the paper a vector notation with $(\cdot,\cdot)$ standing for the scalar product in whatever space the argument may lie. E.g. the expression $(y,\xi_i)$ stands for $\sum_{\mu=1}^M \xi_i^\mu y_\mu$, etc.

We define random maps $m_N^\mu[\omega] : \mathcal{S}_N \to [-1, 1]$ through[1]

$$m_N^\mu[\omega](\sigma) \equiv \frac{1}{N}\sum_{i=1}^N \xi_i^\mu[\omega]\sigma_i \tag{1.1}$$

Naturally, these maps 'compare' the configuration $\sigma$ globally to the random configuration $\xi^\mu[\omega]$. A Hamiltonian is now defined as the simplest negative function of these variables, namely

$$\begin{aligned}H_N[\omega](\sigma) &\equiv -\frac{N}{2}\sum_{\mu=1}^{M(N)} \left(m_N^\mu[\omega](\sigma)\right)^2 \\ &= -\frac{N}{2}\|m_N[\omega](\sigma)\|_2^2\end{aligned} \tag{1.2}$$

where $M(N)$ is some, generally increasing, function that crucially influences the properties of the model. $\|\cdot\|_2$ denotes the $\ell_2$-norm in $I\!\!R^M$, and the vector $m_N[\omega](\sigma)$ is always understood to be $M(N)$-dimensional.

Through this Hamiltonian we define in a natural way finite volume Gibbs measures on $\mathcal{S}_N$ via

$$\mu_{N,\beta}[\omega](\sigma) \equiv \frac{1}{Z_{N,\beta}[\omega]}e^{-\beta H_N[\omega](\sigma)} \tag{1.3}$$

and the induced distribution of the overlap parameters

$$\mathcal{Q}_{N,\beta}[\omega] \equiv \mu_{N,\beta}[\omega] \circ m_N[\omega]^{-1} \tag{1.4}$$

The normalizing factor $Z_{N,\beta}[\omega]$, given by

$$Z_{N,\beta}[\omega] \equiv 2^{-N}\sum_{\sigma \in \mathcal{S}_N} e^{-\beta H_N[\omega](\sigma)} \equiv I\!\!E_\sigma e^{-\beta H_N[\omega](\sigma)} \tag{1.5}$$

is called the partition function. We will frequently consider the non-normalized probabilities that $m_N(\sigma)$ lies in a ball in $I\!\!R^M$ of radius $\rho$ centered at $m$,

$$Z_{N,\beta,\rho}[\omega](m) \equiv I\!\!E_\sigma e^{-\beta H_N[\omega](\sigma)}\,1\!\!1_{\{\|m_N[\omega](\sigma) - m\|_2 \leq \rho\}} \tag{1.6}$$

---

[1] We will make the dependence of random quantities on the random parameter $\omega$ explicit by an added $[\omega]$ whenever we want to stress it. Otherwise, we will frequently drop the reference to $\omega$ to simplify the notation.

We are interested in the exponential asymptotics of these quantities, i.e. in the behaviour of the functions

$$f_{N,\beta,\rho}[\omega](m) \equiv -\frac{1}{\beta N} \ln Z_{N,\beta,\rho}[\omega](m) \tag{1.7}$$

and in particular in the location of the critical points of these functions when $N$ tends to infinity, since these determine not only the asymptotic properties of the Gibbs measures, but also the long-time features of a stochastic dynamics (the so-called "retrieval dynamics") chosen such that the Gibbs measures are their equilibrium distribution.

A study of these functions has been undertaken in a number of previous papers, using either the so-called Hubbard-Stratonovich transformation [FP,K,BGP1], or standard large deviation estimates [BG2]. In the Hubbard-Stratonovich approach, one considers instead of the measure $\mathcal{Q}_{N,\beta}$ itself its convolution with a Gaussian measure on $\mathbb{R}^M$ of mean zero and variance $(\beta N)^{-1} \mathbb{1}$ (where $\mathbb{1}$ is the identity matrix). The resulting measure $\widetilde{\mathcal{Q}}_{N,\beta}$ is absolutely continuous and has a density proportional to

$$\exp\left(-\Phi_{N,\beta}[\omega](z)\right) \tag{1.8}$$

with respect to $M$-dimensional Lebesgue measure. The function $\Phi_{N,\beta}(z)$ can be computed explicitly and is given by

$$\Phi_{N,\beta}[\omega](z) = \frac{1}{2}\|z\|_2^2 - \frac{1}{\beta N}\sum_{i=1}^{N} \ln \cosh \beta(\xi_i, z) \tag{1.9}$$

The results obtained in [BGP1,BGP2] on the concentration of the limiting Gibbs measures were based on an analysis of the location of the absolute minima of the function $\Phi_{N,\beta}$. One may notice that the measures $\widetilde{\mathcal{Q}}_{N,\beta}$ and $\mathcal{Q}_{N,\beta}$ are related by a convolution with a measure that is, asymptotically as $N \uparrow \infty$, concentrated sharply on a sphere of radius $\sqrt{\alpha/\beta}$.

This allows to recover localization properties of the measure $\mathcal{Q}_{N,\beta}$ up to that precision from those of $\widetilde{\mathcal{Q}}_{N,\beta}$. An alternative approach using standard large deviation estimates can also be used (see [BG2]) and reveals that as far as the analysis of the critical points of $f_{N,\beta,\rho}(m)$ is concerned, this also boils down to the study of the same function $\Phi_{N,\beta}$. Notably, the lower large deviation estimates can be obtained only for $\rho \geq \sqrt{2\alpha}$, so that in this way virtually the same precision on localization properties is obtained, and both approaches seem practically equivalent and may be used alternatively according to what appears more convenient in a given situation.

We see that in any case, further progress relies on better estimates on the behaviour of this function and it is the purpose of the present paper to provide a considerably more precise analysis of them then those given in [BGP1]. In particular we get (up to constants) the conjectured behaviour of the critical temperature as a function of $\alpha$, for $\alpha$ small. Let us formulate our main results. We denote here and in the sequel by $m^*(\beta)$ the largest solution of the equation $m = \tanh(\beta m)$. Note that $m^*(\beta)$ is strictly positive for all $\beta > 1$, $\lim_{\beta\uparrow\infty} m^*(\beta) = 1$, and $\lim_{\beta\downarrow 1} \frac{(m^*(\beta))^2}{3(\beta-1)} = 1$. Let us denote by $B_\rho(x)$ the ball of radius $\rho$ centered at $x$ in $I\!\!R^M$. We denote by $e^\mu$ the $\mu$-th unit vector in $I\!\!R^M$. We will see that the relevant small parameter in our problem is always the ratio between $\sqrt{\alpha}$ and $(m^*(\beta))^2$. We will therefor use the general convention to set $\sqrt{\alpha} = \gamma(m^*(\beta))^2$ and we will treat $\gamma$ as our small parameter. Our main results can then be summarized in the following theorems (which however do not contain all the precise estimates on constants that can be found in the later sections).

**Theorem 1:** There exists $\gamma_a > 0$ such that for all $\beta > 1$ for $\alpha \leq \gamma_a^2(m^*(\beta))^4$ there exists constants $c_0 < 1/2$, $c_1 > 0$ such that $I\!\!P$-almost surely for all but a finite number of indices $N$ for all $m \in \left\{\bigcup_{(\mu,s)} B_{c_0\gamma m^*}(sm^*e^\mu)\right\}^c$,

$$\Phi_{N,\beta}[\omega](m) - \Phi_{N,\beta}[\omega](m^*e^1) \geq c_1(m^*)^2 \inf_{(\mu,s)} \|m - sm^*e^\mu\|_2^2 \tag{1.10}$$

**Theorem 2:** Let $z^{(\mu)} \in I\!\!R^{M(N)}$ denote the random vector whose $\nu$-th component is $z_\nu^{(\mu)} \equiv \frac{1}{N}\sum_{i=1}^N \xi_i^\mu \xi_i^\nu$, if $\nu \neq \mu$ and $z_\mu^{(\mu)} \equiv 0$. There exists $\gamma_c > 0$ such that for all $\beta > 1$ for $\alpha \leq \gamma_c^2(m^*(\beta))^4$ there exists strictly positive constants $c_2, c_3$ such that $I\!\!P$-almost surely for all but finitely many $N$, for all $v$ such that $\|v\|_2 \leq c_3 m^*$ and

$$\left\|v - z^{(\mu)}\frac{m^*(\beta)}{1 - \beta(1 - (m^*)^2)}\right\|_2 \geq c_2 m^*(\beta)\gamma^{3/2}\sqrt{|\ln\gamma|} \tag{1.11}$$

then

$$\Phi_{N,\beta}[\omega](m^*e^\mu + v) > \inf_{\|v\|_2 \leq c_3 m^*} \Phi_{N,\beta}[\omega](m^*e^\mu + v) \tag{1.12}$$

We obtain bounds on the various constants in the different asymptotic regimes in the course of the proofs. Our bound on the constant $\gamma_c$ will be considerably larger (of order 0.04 for $\beta$ large) than the one for $\gamma_a$ (of order $10^{-4}$), in accordance with the general expectation that the local minima corresponding to the patterns persist for values of $\alpha$ where they are no longer the absolute minima. Let us remark that a very similar analysis could also be carried out to prove the existence of further local minima associated to so-called "mixed states" (see e.g. [N]), but we leave this to the interested reader.

As a consequence of the previous theorems and the estimates entering their proofs we get the following theorem on the Gibbs measures.

**Theorem 3:** For all $\beta > 1$ and $\alpha \leq \gamma_a^2(m^*(\beta))^4$ there exists a constant $c_5 < \frac{1}{2\gamma_a}$ such that

$$\lim_{N\uparrow\infty} \mu_{N,\beta}[\omega]\left(\left\{\bigcup_{(\mu,s)} B_{c_5\gamma m^*}(se^\mu m^*)\right\}\right) = 1, \quad I\!\!P - a.s. \tag{1.13}$$

Moreover, for any pair of indices $\mu, \nu$,

$$\lim_{N \uparrow \infty} \frac{1}{N} \ln \left[ \frac{\mu_{N,\beta}[\omega] \left( B_{c_5 \gamma m^*}(e^\mu m^*) \right)}{\mu_{N,\beta}[\omega] \left( B_{c_5 \gamma m^*}(e^\nu m^*) \right)} \right] = 0 \quad I\!P - a.s. \tag{1.14}$$

**Remark:** Theorem 3 sharpens the results of [BGP1] and [BGP2]. (1.14) guarantees that limiting measures concentrated on a single ball can be constructed by applying an magnetic field aligned on one of the patterns whose strength can be taken to zero after the limit $N \uparrow \infty$ is taken. See [BGP1] for a general discussion on limiting Gibbs measures. In a recent note [T2] Talagrand has announced an estimate similar to (1.13) under some additional restrictions on $\beta$.

The remainder of this paper is structured as follows. The next section introduces a new very sharp bound on the behaviour of the maximal eigenvalue of the random matrix $\xi^t \xi$. While we believe that this result has some interest in itself in that it provides considerably sharper bounds than were previously available (the sharpest ones, to our knowledge, being due to Shcherbina and Tirozzi [ST] were of the order $\exp(-N^{2/3})$ only), this introduces some of the basic 'new' techniques in a rather simple situation and can thus be seen as a warm up for what will follow. In section 3 we improve the estimates of [BGP1] by locating more precisely the absolute minima of $\Phi_{N,\beta}$ for very small $\alpha$. Section 4 is the central part of this work. Here we control the precise location of the local minima corresponding to the patterns and control the behaviour of $\Phi_{N,\beta}$ near them. The main difficulty we have to overcome here is that the function $\Phi_{N,\beta}$ is random. The usual way to get precise estimates on a function near its minima is to use a Taylor expansion. Due to the randomness, there can be no uniform control over the remainder terms, but we have to deal with the probabilities of large excursions. To estimate those, we need to control suprema of certain random processes that are indexed by continuous parameters taking values in high-dimensional sets. In this analysis we invoke techniques introduced in the analysis of the regularity of random processes in Banach spaces (see [LT]). This rather long section is subdivided into three subsections: In part 1 we prove the uniform upper and lower bounds on $\Phi$. In part 2 these are used to localize the position of the minima. Here we also prove the local convexity of $\Phi$. In part 3 we localize the value of the unique macroscopic component of the position of the minima and show that in the limit $\beta \uparrow \infty$ it differs from one by an term proportional to $\exp(-1/2\alpha)$. In Section 5 we apply the previous estimates to prove Theorem 3. An appendix contains the proof of a technical lemma needed in Section 4.3.

**Acknowledgements:** We thank Michel Talagrand for sending us a copy of [T1] through which we learned about Theorem 2.5. We are grateful to Barbara Gentz for helpful comments on earlier drafts of this paper. A.B. thanks Dmitry Ioffe for useful discussions on the proof of Lemma 4.18.

## 2. An exponential bound on random matrix norms

As a technical warm-up for what is to come, as well as a basic input for the remainder, we will show how techniques of the types used in the analysis of random processes (for an exposition see e.g. [LT]) and concentration of measure estimates (we refer explicitly to the recent paper [T1] by M.Talagrand) can be used to get exponential bounds on the maximal eigenvalues of random matrices that are relevant for our analysis. Note that subexponential bounds have been known for a long time and were generally used in our previous analysis [ST,K,BG2,BGP1].

We are interested in the matrix $A_N \equiv \frac{\xi^t \xi}{N}$. (To simplify notation, we will frequently drop the index $N$ and write $A$ for the matrix $A_N$ in the generic dimension $N$). We begin with the simplest a-priori estimate on the corresponding quadratic form:

**Lemma 2.1:** *For any non zero $x \in \mathbb{R}^M$ and for all $c > 0$*

$$I\!P\left[(x, A_N x) \geq (1+c)\|x\|_2^2\right] \leq \exp\left\{-\frac{N}{2}\left(c - \ln(1+c)\right)\right\} \tag{2.1}$$

**Proof:** We simply use the exponential Chebeychev inequality and the Hubbard Stratonovich transformation [HS] to see that

$$\begin{aligned}
P\left[(x, Ax) \geq (1+c)\|x\|_2^2\right] &= I\!P\left[\frac{1}{N}\sum_{i=1}^N \left(\frac{1}{\|x\|_2}\sum_{\mu=1}^M x_\mu \xi_i^\mu\right)^2 > 1+c\right] \\
&\leq \inf_{0 < t < 1/2} e^{-t(1+c)N}\left[I\!E e^{t\left(\frac{1}{\|x\|_2}\sum_{\mu=1}^M x_\mu \xi_1^\mu\right)^2}\right]^N \\
&\leq \inf_{0 < t < 1/2} e^{-t(1+c)N}\left[\frac{1}{\sqrt{1-2t}}\right]^N
\end{aligned} \tag{2.2}$$

Now the infimum over $t$ in the last line of (2.2) is taken on for $t = \frac{1}{2}\frac{c}{1+c}$ and inserting this value in (2.2) yields (2.1). ◇

Let us now introduce a family of grids $\mathcal{W}_{M,r}$ in $\mathbb{R}^M$ with spacing $\frac{r}{\sqrt{M}}$. We denote by $\mathcal{W}_{M,r}(\rho)$ the set of points $x \in \mathcal{W}_{M,r}$ such that $\|x\|_2 \leq \rho$. We have

**Lemma 2.2:** *Let $B_r(x)$ denote the ball of radius $r$ centered at $x$. Then*

*(i) $\cup_{x \in \mathcal{W}_{M,r}(\rho)} B_r(x) \supset B_\rho(0)$*

*(ii) $|\mathcal{W}_{M,r}(\rho)| \leq e^{M(\ln + \frac{\rho}{r} + c)}$, for some constant $c < 1$.*

**Proof:** Statement (i) follows since the length of the diagonal in a $M$-dimensional cube of side length $\frac{r}{\sqrt{M}}$ equals $r$. Statement (ii) reflects the fact that the volume of a ball of radius $\rho$ in $\mathbb{R}^M$ is $\left(2\rho^M \pi^M\right)/(M\Gamma(M/2))$. ◇

We will control the norm of the matrices $A$ by using the definition of the matrix norm

$$\|A\| \equiv \sup_{x: \|x\|_2 = 1} (x, Ax) \tag{2.3}$$

To estimate the probabilities of suprema over continuous sets of random variables, we will employ a technique used by Ledoux and Talagrand for instance in their textbook [LT]. To this end we fix a number $a < 1$ to be chosen later and chose a sequence $r_n = a^n$. Then any $x$ with norm one can be written in some (possibly non-unique) way as

$$x = \sum_{n=1}^{n^*+1} x(n), \tag{2.4}$$

where $x(n) \in \mathcal{W}_{M, r_n}(r_{n-1})$ for $n \leq n^*$ and $\|x(n^* + 1)\|_2 \leq r_{n^*}$. We will abbreviate for simplicity $\mathcal{W}(n) \equiv \mathcal{W}_{M, r_n}(r_{n-1})$. This gives that

$$\sup_{x: \|x\|_2=1} (x, Ax) = \sup_{x(1) \in \mathcal{W}(1)} \ldots \sup_{x(n^*) \in \mathcal{W}(n^*)} \sup_{x(n^*+1): \|x(n^*+1)\|_2 \leq r_{n^*}} \left( \sum_n x(n), A \sum_n x(n) \right) \tag{2.5}$$

To make good use of this formula, the following elementary lemma is of great help:

**Lemma 2.3:** *Let $b_n$, $n \geq 1$, be any absolutely summable sequence of real numbers. Then, for all $q^2 > 0$,*

$$\left( \sum_{n=1}^{n^*+1} b_n \right)^2 \leq (1 + q^2) \sum_{n=1}^{n^*} (1 + q^{-2})^{n-1} b_n^2 + (1 + q^{-2})^{n^*} b_{n^*+1}^2 \tag{2.6}$$

Of course this formula is useful only if $b_n^2 (1 + q^{-2})^{n-1}$ is summable.

**Proof:** The proof of this lemma follows by induction from the elementary observation that for all $q^2 > 0$, first

$$2x = 2qq^{-1}x = q^2 + q^{-2}x^2 - (q - q^{-1}x)^2 \leq q^2 + q^{-2}x^2 \tag{2.7}$$

and whence

$$\begin{aligned}(b + c)^2 &= b^2 \left( 1 + 2\frac{c}{b} + \frac{c^2}{b^2} \right) \\ &\leq b^2 + c^2 + q^2 b^2 + q^{-2} c^2 = b^2 (1 + q^2) + c^2 (1 + q^{-2}) \end{aligned} \tag{2.8}$$

◇

Lemma 2.3 allows us to write, for $q$ to be chosen later, that

$$\sup_{x \in \mathbb{R}^M: \|x\|_2=1} (x, Ax) \leq (1 + q^2) \sum_{n=1}^{n^*} (1 + q^{-2})^{n-1} \sup_{x(n) \in \mathcal{W}(n)} (x(n), Ax(n)) \\ + (1 + q^{-2})^{n^*} \sup_{x(n^*+1): \|x(n^*+1)\|_2 \leq r_{n^*}} (x(n^* + 1), Ax(n^* + 1)) \tag{2.9}$$

But combining (2.1) with (ii) of Lemma 2.2, we get that

$$I\!P\left[\sup_{x(n)\in\mathcal{W}(n)}(x(n),Ax(n))\geq (1+c)a^{2(n-1)}\right]\leq e^{M(|\ln a|+1)-Ng(c)} \quad (2.10)$$

where we have set

$$g(c) \equiv \tfrac{1}{2}\left\{c - \ln(1+c)\right\} \quad (2.11)$$

Therefore,

$$I\!P\left[\sum_{n=1}^{n^*}(1+q^2)(1+q^{-2})^{n-1}\sup_{x(n)\in\mathcal{W}(n)}(x(n),Ax(n))\geq(1+c)(1+q^2)\sum_{n=1}^{n^*}(1+q^{-2})^{n-1}a^{2(n-1)}\right]$$
$$\leq \sum_{n=1}^{n^*}I\!P\left[\sup_{x(n)\in\mathcal{W}(n)}(x(n),Ax(n))\geq(1+c)\|x(n)\|_2^2\right]$$
$$\leq n^* e^{M(|\ln_+ a|+1)-Ng(c)}$$
$$(2.12)$$

On the other hand, it is a trivial matter to see that uniformly,

$$(1+q^{-2})^{n^*}(x(n^*+1),Ax(n^*+1))\leq M(1+q^{-2})^{n^*}\|x(n^*+1)\|_2^2 \leq M((1+q^{-2})a^2)^{n^*} \quad (2.13)$$

We thus obtain, combining our estimates,

$$I\!P\left[\sup_{x:\|x\|_2=1}(x,Ax)\geq (1+c)\left\{\frac{(1+q^2)}{1-(1+q^{-2})a^2}+M((1+q^{-2})a^2)^{n^*}\right\}\right]\leq n^* e^{M(|\ln a|+1)-Ng(c)} \quad (2.14)$$

Of course the constants $q$ and $a$ have been assumed to satisfy $(1+q^{-2})a^2 < 1$. It remains now to choose these constants as well as $n^*$. Without attempting a strict optimization, a reasonable choice turns out to be, for $\sqrt{\alpha}\leq 1/2$,

$$q^2 = a = \sqrt{\alpha} \quad (2.15)$$

With this choice, the remainder term (2.13) is bounded by $1/N$ if $n^* = \ln(MN)/|\ln\left(\tfrac{3}{4}\right)|$. If we moreover set $c \equiv g^{-1}\left(\alpha(|\ln\alpha|/2+1)+\epsilon\right)$, $\epsilon > 0$, (2.14) finally gives

$$I\!P\left[\sup_{x:\|x\|=1}(x,Ax)\geq\left\{\frac{1+\sqrt{\alpha}}{1-\alpha-\sqrt{\alpha}}+\frac{1}{N}\right\}\left(1+g^{-1}\left(\alpha(|\ln\alpha|/2+1)+\epsilon\right)\right)\right]\leq\frac{\ln(MN)}{|\ln\left(\tfrac{3}{4}\right)|}e^{-\epsilon N} \quad (2.16)$$

This bound is not very good to determine the true norm of $A$, but it gives very good estimates on probabilities of very large excesses. We will now bootstrap this result with the help of a general 'concentration of measure' theorem of M. Talagrand [T1]. To this end we need the following properties of the norm of $A$ as a function of $\xi$.

**Lemma 2.4:** Set $\lambda_N(\omega)\equiv\sup_{x:\|x\|=1}(x,A_N[\omega]x)$. Then

*(i) The function $\lambda_N(\omega)$ is a convex function of the random variables $\xi(\omega)$.*

*(ii) $\lambda_N(\omega)$ satisfies the Lipshitz bound*

$$|\lambda_N(\omega) - \lambda_N(\omega')| \leq \frac{\sqrt{2}}{\sqrt{N}} \sqrt{\lambda_N(\omega) + \lambda_N(\omega')} \|\xi(\omega) - \xi(\omega')\|_2 \qquad (2.17)$$

**Proof:** To prove (i), note that $(x, Ax) = \frac{1}{N} \sum_i (\xi_i, x)^2$ is a convex function of $\xi$ for fixed $x$. But the supremum of a family of convex functions is again convex.

To prove (ii), note first that

$$\left| \sup_{x: \|x\|=1} (x, A(\omega)x) - \sup_{x: \|x\|=1} (x, A(\omega')x) \right| \leq \sup_{x: \|x\|=1} |(x, A(\omega)x) - (x, A(\omega')x)| \qquad (2.18)$$

But

$$|(x, A(\omega)x) - (x, A(\omega')x)| = \left| \frac{1}{N} \sum_i \left( (\xi_i(\omega), x)^2 - (\xi_i(\omega'), x)^2 \right) \right|$$

$$= \left| \frac{1}{N} \sum_i (\xi_i(\omega) + \xi_i(\omega'), x)(\xi_i(\omega) - \xi_i(\omega'), x) \right| \qquad (2.19)$$

$$\leq \sqrt{\frac{1}{N} \sum_i (\xi_i(\omega) + \xi_i(\omega'), x)^2} \sqrt{\frac{1}{N} \sum_i (\xi_i(\omega) - \xi_i(\omega'), x)^2}$$

Now

$$\frac{1}{N} \sum_i (\xi_i(\omega) + \xi_i(\omega'), x)^2 \leq \frac{2}{N} \sum_i (\xi_i(\omega), x)^2 + \frac{2}{N} \sum_i (\xi_i(\omega'), x)^2 \qquad (2.20)$$

while

$$\frac{1}{N} \sum_i (\xi_i(\omega) - \xi_i(\omega'), x)^2 \leq \frac{1}{N} \sum_i \sum_\mu (\xi_i(\omega) - \xi_i(\omega'))^2 \sum_\mu x_\mu^2 = \frac{1}{N} \|\xi(\omega) - \xi(\omega')\|_2^2 \|x\|_2^2 \qquad (2.21)$$

from which (2.19) follows. $\diamond$

**Theorem 2.5:** ([T1]) *Let $f$ be a real valued function defined on $[-1, 1]^N$. Assume that for each real number $a$, the set $\{f \leq a\}$ is convex. Suppose that on a convex set $B \subset [-1, 1]^N$ the restriction of $f$ to $B$ satisfies for all $x, y \in B$*

$$|f(x) - f(y)| \leq l_B \|x - y\|_2 \qquad (2.22)$$

*for some constant $l_B > 0$. Let $h$ denote the random variable $h = f(X_1, \ldots, X_N)$. Then, if $M_f$ is a median of $h$, for all $t > 0$,*

$$I\!P[|h - M_f| \geq t] \leq 4b + \frac{4}{1 - 2b} \exp\left( -\frac{t^2}{16 l_B^2} \right) \qquad (2.23)$$

*where b denotes the probability of the complement of the set B.*

We see that due to Lemma 2.4 we are exactly in the situation where we may apply this theorem with $h$ being the norm of $A$.

This gives us the following

**Theorem 2.6:** *Assume that $\alpha \leq 1/4$. Then there exists a constant $K = K(\alpha) < \infty$ such that for all $x \leq 1$*

$$I\!P\left[|\|A\| - I\!E\|A\||  \geq x\right] \leq K e^{-\frac{Nx^2}{K}} \tag{2.24}$$

*The same result holds for $A$ replaced by $A - 1\!\!1$.*

**Remark:** From Theorem 2.5 we get an exponential estimate on $|\|A\| - M_{\|A\|}|$. But it is easy to see that this together with (2.19) in turn implies the exponential estimate (2.24) (with slightly modified constants).

From the known standard estimates on the eigenvalues of $A$ (the first reference to our knowledge is [Ge]) we know that the median of $\|A\|$ equals $(1+\sqrt{\alpha})^2$ and that of $\|A - 1\!\!1\|$ equals $2\sqrt{\alpha}+\alpha$, up to corrections that tend to zero with $N$ rapidly.

**Proof:** Theorem 2.6 is a direct consequence of Lemma 2.4 and Theorem 2.5, together with the estimate (2.16), used for some suitable small value of $\epsilon$. Since Lemma 2.4 holds also for the norm of $A - 1\!\!1$, we get the same estimate for the norm of that matrix. The constant $K(\alpha)$ can be estimated more precisely from our bounds, but its value will be of no particular importance for the rest of this paper. $\diamond$

Theorem 2.6 will be used heavily in the remainder of this paper. We introduce, for future reference the sets

$$\Omega_1(N) \equiv \left\{\omega \in \Omega \big| \|A_N[\omega] - 1\!\!1\| \leq r_N(\alpha)\right\} \tag{2.25}$$

and

$$\Omega_1 \equiv \cup_{N_0 \geq 1} \cap_{N \geq N_0} \Omega_1(N) \tag{2.26}$$

where $r_N(\alpha) \equiv 2\sqrt{2} + \alpha + \epsilon$, for some arbitrarily small $\epsilon$ (one may also take $\epsilon$ that decrease with $N$, e.g. $\epsilon = C\sqrt{\ln N/N}$). Then one has that $I\!P[\Omega_1(N)] \geq 1 - K\exp(-N\epsilon^2/K)$ and $I\!P[\Omega_1] = 1$.

# 3. Global minima

In this section we determine a regime in the $\alpha, \beta$ plane for which global minima away from the Mattis states can be excluded. This will provide a more transparent proof and better estimates on the parameters than previously obtained in [BGP1]. In particular, it will yield the correct asymptotic behaviour of the maximal allowed $\alpha$ for $\beta \downarrow 1$ which agrees (up to constants) with the findings from replica methods [AGS].

We first introduce the following subsets of $I\!R^M$:

$$\Gamma_\epsilon = \{m \mid |\|m\|_2 - m^*| > \epsilon m^*\} \tag{3.1}$$

and

$$D_{\rho,\epsilon} = \Gamma_\epsilon^c \bigcap \left\{ \bigcup_{(\mu,s)} B_\rho(sm^* e^\mu) \right\}^c \tag{3.2}$$

where the union runs over all $(\mu, s) \in \{1, \ldots, M\} \times \{-1, 1\}$ and where $B_\rho(m)$ denotes the ball of radius $\rho$ centered at $m$.

The central result of this section is the following theorem.

**Theorem 3.1:** *There exists strictly positive constants $\gamma_a, c_1, c_2, \bar{c}$ such that for all $0 \leq \alpha \leq \gamma_a^2 (m^*(\beta))^4$ there exists a set $\tilde{\Omega} \subset \Omega$ with $I\!P[\tilde{\Omega}^c] \leq e^{-c_1 N}$ and a constant $0 < c_4 < \frac{1}{2}$, such that for all $\omega \in \tilde{\Omega}$ the function $\Phi_{N,\beta}[\omega](m)$ satisfies the following:*

*(i) For all $m \in \Gamma_{1/35}$*

$$\Phi_{N,\beta}[\omega](m) - \phi(m^*) \geq \tfrac{1}{2}\bar{c}\,(m^*)^2 \left(\|m\|_2 - m^*\right)^2 \tag{3.3}$$

*and*

*(ii) For all $m \in D_{c_4 m^*, 1/35}$,*

$$\Phi_{N,\beta}[\omega](m) - \phi(m^*) \geq c_2\,(m^*)^4 \tag{3.4}$$

*In particular, all absolute minima of $\Phi$ lie in the union of the balls $B_{c_4 m^*}(\pm m^* e^\mu)$.*

**Remark:** The proof of this theorem provides estimates on the constants $c_i$ that we have not tried to optimize. The interested reader is invited to do this. The relation between the critical $\alpha$ and $\beta - 1$ show however the expected correct power-law behaviour. Note that asymptotically, as $\beta \downarrow 1$, $m^*(\beta)^2 \approx 3(\beta - 1)$.

**Proof:** Let us first give a brief outline of the proof. We will treat separately the regions $\Gamma_\epsilon$, $D_{\epsilon, 1/2}$ and the balls $B_{1/2}(sm^* e^\mu)$. On the first two sets we will use that on the set $\Omega_1$ defined in (2.26),

$$\Phi(m) - \phi(m^*) \geq -\tfrac{r(\alpha)}{2}\|m\|_2^2 + \tfrac{1}{N}\sum_i \left(\phi\left(\langle\xi_i, m\rangle\right) - \phi(m^*)\right) \tag{3.5}$$

and prove a suitable lower bound on

$$\frac{1}{N} \sum_i \left( \phi\left((\xi_i, m)\right) - \phi(m^*) \right) \tag{3.6}$$

To treat the balls $B_{1/2}(sm^* e^\mu)$ we will, performing the change of variables $m = sm^* e^\mu + v$, use that on $\Omega_1$

$$\Phi(m) - \phi(m^*) \geq -\frac{r(\alpha)}{2} \|v\|_2^2 - m^* r(\alpha) \|v\|_2 + \frac{1}{N} \sum_i \left( \phi\left((\xi_i, m)\right) - \phi(m^*) \right) \tag{3.7}$$

to show that

$$\frac{1}{N} \sum_i \left( \phi\left((\xi_i, m)\right) - \phi(m^*) \right) \geq c'(\beta) \|v\|_2^2 \tag{3.8}$$

$\|v\|_2 > \frac{m^* r(\alpha)}{c'(\beta) - r(\alpha)/2}$ then guarantees $\Phi(m) > \phi(m^*)$. Of course this requires $c'(\beta) > r(\alpha)/2$.

We start with the following preparatory lemmas

**Lemma 3.2:** *Let*

$$\hat{c}(\beta) \equiv \frac{\ln \cosh(\beta m^*)}{\beta (m^*)^2} - \frac{1}{2} \tag{3.9}$$

*Then for all $\beta > 1$ and for all $z$*

$$\phi(z) - \phi(m^*) \geq \hat{c}(\beta)(|z| - m^*)^2 \tag{3.10}$$

*Moreover $\hat{c}(\beta)$ tends to $\frac{1}{2}$ as $\beta \uparrow \infty$, and behaves like $\frac{1}{6}(m^*(\beta))^2$, as $\beta \downarrow 1$.*

**Proof:** Notice that the function $\phi(z)$ is symmetric and has the property that $z \phi'''(z) \geq 0$. Considering only the positive branch, we see that the constant $\hat{c}(\beta)$ was chosen such that equality holds in (3.10) at the points $0$ and $m^*$. To show that this implies that the quadratic function is a lower bound is an exercise in elementary calculus.

The asymptotic behaviour of $\hat{c}(\beta)$ follows form the fact that for small argument, $\ln \cosh x \approx x^2/2$, while for large arguments $\ln \cosh x \approx |x|$. $\Diamond$

**Lemma 3.3:** *On $\Omega_1$,*

*(i) If $\|m\|_2 \sqrt{1 - r(\alpha)} > m^*$, then*

$$\frac{1}{N} \sum_{i=1}^N \phi\left((\xi_i, m)\right) - \phi(m^*) \geq \hat{c}(\beta) \left( \|m\|_2 \sqrt{1 - r(\alpha)} - m^* \right)^2 \tag{3.11}$$

*and*

(ii) if $\|m\|_2 \sqrt{1 + r(\alpha)} < m^*$, then
$$\frac{1}{N} \sum_{i=1}^{N} \phi\left((\xi_i, m)\right) - \phi(m^*) \geq \hat{c}(\beta) \left(\|m\|_2 \sqrt{1 + r(\alpha)} - m^*\right)^2 \tag{3.12}$$

**Proof:** Using Lemma 3.2, we see that
$$\frac{1}{N} \sum_{i=1}^{N} \phi\left((\xi_i, m)\right) - \phi(m^*) \geq \hat{c}(\beta) \frac{1}{N} \sum_{i=1}^{N} \left[(\xi_i, m)^2 - m^* |(\xi_i, m)| + (m^*)^2\right]$$
$$= \hat{c}(\beta) \left[\frac{\|\xi m\|_2}{\sqrt{N}} - m^*\right]^2 \tag{3.13}$$
where we used the Schwarz inequality. From here the lemma follows by using the bounds on the norms of the random matrices $\xi^t \xi / N$ established in Section 2. $\Diamond$

**Corollary 3.4:** There exists a constant $c_1 > 0$ such that if $\sqrt{\alpha} \leq c_1 (m^*)^2$, then there exists $\epsilon = \epsilon(\alpha) \approx \sqrt{2\sqrt{\alpha}/\hat{c}(\beta)}$ such that if $m \in \Gamma_\epsilon$ then for all $\omega \in \Omega_1$,
$$\Phi_{N,\beta}[\omega](m) - \phi(m^*) \geq \tfrac{1}{2} \hat{c}(\beta)(\|m\|_2 - m^*)^2 \tag{3.14}$$

**Proof:** This follows from the preceeding lemma by elementary algebra. $\Diamond$

This concludes our treatment of the region $\Gamma_\epsilon$. The case of the region $D_{\epsilon,1/2}$ and the balls $B_{1/2}(sm^* e^\mu)$ will be more involved. In particular, we will get a priori only probabilistic versions of the analogs of Lemma 3.3, and thus we will have to estimate probabilities of suprema over $m$ of our functions $\phi(m)$. Our first observation is thus that the function $\Phi(m)$ is Lipshitz continuous on $\Omega_1$ which will allow us to reduce the problem to an estimate of a lattice supremum. We have

**Lemma 3.5:** For all $\omega \in \Omega_1$ and for all $\beta$,
$$|\Phi(m) - \Phi(m')| \leq \left(\frac{\|m\|_2 + \|m'\|_2}{2} + \sqrt{1 + r(\alpha)}\right) \|m - m'\|_2 \tag{3.15}$$

**Proof:** The proof of this lemma consists just in some applications of the Schwarz inequality. Note that of course
$$\|m\|_2^2 - \|m'\|_2^2 = (m + m', m - m') \leq \|m + m'\|_2 \|m - m'\|_2 \tag{3.16}$$
On the other hand it follows from the mean value Theorem in $\mathbb{R}^M$ that for some $0 < \theta < 1$,
$$\frac{1}{\beta N} \sum_i \left(\ln \cosh(\beta(\xi_i, m)) - \ln \cosh(\beta(\xi_i, m'))\right)$$
$$= \frac{1}{N} \sum_i (\xi_i, m - m') \tanh(\beta(\xi_i, m' + \theta(m - m'))) \tag{3.17}$$
$$\leq \|m - m'\|_2 \sqrt{\frac{\|\xi^t \xi\|}{N}} \sqrt{\frac{1}{N} \sum_i \tanh^2(\beta(\xi_i, m' + \theta(m - m')))}$$

Using the inequality $|\tanh x| \leq 1$ and the bound on the norm of $\xi^t \xi/N$ on $\Omega_1$, we arrive at (3.15).
◇

**Remark:** The bound (3.15) is actually quite poor and can be improved considerably, in particular for $m$ and $m'$ near the critical points of $\Phi$ and if $\beta$ is near one. We leave this as an exercise to the reader. We can live with this simple bound on the expense of choosing a smaller lattice spacing, and this does not substantially deteriorate our results.

**Lemma 3.6:** *Let $X_i \geq 0$, $i = 1, \ldots, N$ be positive i.i.d. random variables that satisfy $I\!P[X_i \geq z] \geq q$. Then for all $\zeta \geq 0$,*

$$I\!P\left[\frac{1}{N}\sum_{i=1}^{N} X_i \leq qz(1-\zeta)\right] \leq \exp\left\{-Nq\frac{\zeta^2}{2}\right\} \tag{3.18}$$

**Proof:** By the exponential Markov inequality we have that

$$\begin{aligned} I\!P\left[\frac{1}{N}\sum_{i=1}^{N} X_i \leq qz(1-\zeta)\right] &\leq \inf_{t\geq 0} e^{tqz(1-\zeta)N} \left[I\!E e^{-tX_1}\right]^N \\ &\leq \inf_{t\geq 0} e^{tqz(1-\zeta)N} \left[1 + q(e^{-tz} - 1)\right]^N \end{aligned} \tag{3.19}$$

Choosing $t = \epsilon/z$ and using the inequality

$$\ln\left[1 + q(e^{-\zeta} - 1)\right] \leq -q\zeta + \frac{q\zeta^2}{2} \tag{3.20}$$

one obtains (3.18).◇

Lemma 3.6 will be used together with the following observation.

**Lemma 3.7:** *Let $1 \leq t \leq M$ be a fixed integer. For any $m \in I\!R^M$ set*

$$\begin{aligned} \hat{m} &\equiv (m_1, \ldots, m_t, 0, \ldots, 0) \\ \tilde{m} &\equiv (0, \ldots, 0, m_{t+1}, \ldots, m_M) \end{aligned} \tag{3.21}$$

*Then, for any $0 < d < 1$,*

$$I\!P\left[\phi((\xi_1, m)) - \phi(m^*) \geq \hat{c}(\beta)(m^*)^2 d^2\right] \geq \frac{1}{2} - \frac{1}{4}\frac{\|m\|_2^2 + (\|\hat{m}\|_2 - \|\tilde{m}\|_2)^2}{(1-d)^2(m^*)^2} \tag{3.22}$$

*where $\hat{c}(\beta)$ is the constant from Lemma 3.2.*

**Proof:** Let us put $X = (\hat{m}, \xi_1)$ and $Y = (\tilde{m}, \xi_1)$. Note that $X$ and $Y$ are independent and symmetric random variables. By Lemma 3.2 we have that

$$I\!P\left[\phi((\xi_1, m)) - \phi(m^*) \geq \hat{c}(\beta)(m^*)^2 d^2\right] \geq 1 - I\!P\left[||X + Y| - m^*| \leq dm^*\right] \tag{3.23}$$

Now
$$IP\left[||X+Y|-m^*|\leq dm^*\right]\leq IP\left[|X+Y|\geq (1-d)m^*\right] \qquad (3.24)$$

By the symmetry of $X$ and $Y$,

$$\begin{aligned}IP\left[|X+Y|\geq (1-d)m^*\right] &= \tfrac{1}{2}IP\left[|X+Y|\geq (1-d)m^*\big|X\geq 0, Y\geq 0\right] \\ &+ \tfrac{1}{2}IP\left[|X+Y|\geq (1-d)m^*\big|X\geq 0, Y<0\right] \\ &\leq \tfrac{1}{2}+\tfrac{1}{2}IP\left[|X+Y|\geq (1-d)m^*\big|X\geq 0, Y<0\right]\end{aligned} \qquad (3.25)$$

For the last inequality we finally use the Chebychev inequality. This gives

$$IP\left[|X+Y|\geq (1-d)m^*\big|X\geq 0, Y<0\right]\leq \frac{IEX^2+IEY^2-2IE|X|IE|Y|}{(1-d)^2(m^*)^2} \qquad (3.26)$$

The announced result follows from here by the Khintchine inequality [Sz], which tells us that $IE|X|\geq \|\hat{m}\|_2/\sqrt{2}$. $\diamond$

To make use of this lemma, one has to choose $t$ in the decomposition (3.21) in such a way that $\|\hat{m}\|_2$ and $\|\tilde{m}\|_2$ are as similar as possible. We may suppose without loss of generality that $m_1\geq |m_2|\geq \ldots \geq |m_M|$. Then the conditions $|\|m\|_2-m^*|\leq \epsilon m^*$ and $\|m-e^1 m^*\|_2\geq \tfrac{1}{2}m^*$ imply that

$$m_1 \leq \left(\tfrac{7}{8}+\epsilon+\tfrac{\epsilon^2}{2}\right)m^* \qquad (3.27)$$

Without loosing anything, we can choose $t=1$ as long as

$$m_1 \geq m^*\sqrt{(1-\epsilon)^2-\left(\tfrac{7}{8}+\epsilon+\tfrac{\epsilon^2}{2}\right)^2} \qquad (3.28)$$

This gives the bound

$$\left(\|\hat{m}\|_2-\|\tilde{m}\|_2\right)^2 \leq (m^*)^2\left(\left(\tfrac{7}{8}+\epsilon+\tfrac{\epsilon^2}{2}\right)-\sqrt{(1-\epsilon)^2-\left(\tfrac{7}{8}+\epsilon+\tfrac{\epsilon^2}{2}\right)^2}\right)^2 \equiv g(\epsilon)(m^*)^2 \qquad (3.29)$$

If $m_1$ is smaller than the value given in (3.28), then we must choose $t$ larger. The point here is that we can always find a $t$ for which

$$\left|\|\hat{m}\|_2^2-\|\tilde{m}\|_2^2\right|\leq m_1^2 \qquad (3.30)$$

and this implies, for these values of $m_1$, an even smaller bound on $(\|\hat{m}\|_2-\|\tilde{m}\|_2)^2$ than $g(\epsilon)(m^*)^2$. Combining now (3.22) and (3.29) with Lemma 3.6, we arrive at the bound

$$\begin{aligned}&IP\left[\tfrac{1}{N}\sum_i \phi((\xi_i,m))-\phi(m^*)\leq \hat{c}(\beta)(m^*)^2 d^2\left[\tfrac{1}{2}-\tfrac{(1+\epsilon)^2+g(\epsilon)}{4(1-d)^2}\right](1-\zeta)\right] \\ &\leq \exp\left(-N\left[\tfrac{1}{2}-\tfrac{(1+\epsilon)^2+g(\epsilon)}{4(1-d)^2}\right]\tfrac{\zeta^2}{2}\right)\end{aligned} \qquad (3.31)$$

for arbitrary positive $\zeta$. This bound looks somewhat complicated, and it is most reasonable to make a choice for $\epsilon$ and $d$. Numerically, it turns out that if we fix $\epsilon = \frac{1}{35}$ and $d \approx 0.102$, then (3.31) gives us the desired

**Lemma 3.8:** *For all $m \in D_{m^*/2, 1/35}$ and for any $\zeta \geq 0$*

$$I\!P\left[\frac{1}{N}\sum_i \phi((\xi_i, m)) - \phi(m^*) \leq \frac{\hat{c}(\beta)(m^*)^2(1-\zeta)}{35^2}\right] \leq \exp\left(-N\frac{\zeta^2}{32}\right) \quad (3.32)$$

We are left to treat the case of the balls $B_{1/2}(sm^* e^\mu)$. W.l.g we will consider the ball $B_{1/2}(m^* e^1)$. We will prove:

**Lemma 3.9:** *Assume that $m \in B_{1/2}(m^* e^1)$. Then,*

$$I\!P\left[\frac{1}{N}\sum_i \phi((\xi_i, m)) - \phi(m^*) \leq \tilde{c}(\beta)\left[1 - r(\alpha) - 0.5\right]\|m - e^1 m^*\|_2^2 \Big| \Omega_1\right] \leq \exp\left(-\frac{N}{50}\right) \quad (3.33)$$

where $\tilde{c}(\beta) = \frac{\phi(3m^*/4) - \phi(m^*)}{(7m^*/4)^2} (\approx \frac{1}{430}\beta^4(m^*)^2$, for $\beta$ near $1$).

**Proof:** Like in Lemma 3.2 it is clear that for $z > -\frac{3}{4}m^*$,

$$\phi(z) - \phi(m^*) \geq \tilde{c}(\beta)(z - m^*)^2 \quad (3.34)$$

if $\tilde{c}(\beta)$ is chosen such that the parabola on the right intersects the function on the left at $z = -3m^*/4$. Thus we can use that

$$\frac{1}{N}\sum_i \phi((\xi_i, m)) - \phi(m^*) \geq \tilde{c}(\beta)\frac{1}{N}\sum_i \left((\xi_i, m) - \xi_i^1 m^*\right)^2$$
$$- \tilde{c}(\beta)\frac{1}{N}\sum_i 1\!\!1_{\{\xi_i^1(\xi_i, m) < -3m^*/4\}}\left(\xi_i^1(\xi_i, m) - m^*\right)^2 \quad (3.35)$$

But since on $\Omega_1$, $\|A - 1\!\!1\| \leq r(\alpha)$,

$$\frac{1}{N}\sum_i \left((\xi_i, m) - \xi_i^1 m^*\right)^2 = \left((m - m^* e^1), \frac{\xi^t \xi}{N}(m - m^* e^1)\right)$$
$$\geq \|m - m^* e^1\|_2 \left(1 - \left\|\frac{\xi^t \xi}{N} - 1\!\!1\right\|\right) \quad (3.36)$$
$$\geq \|m - m^* e^1\|_2^2 (1 - r(\alpha))$$

So that all we have to estimate is

$$I\!P\left[\frac{1}{N}\sum_i 1\!\!1_{\{(\hat{\xi}_i, \hat{m}) < -\frac{7}{4}m^* - v\}}\left((\hat{\xi}_i, \hat{m}) + v\right)^2 > x\right] \quad (3.37)$$

where we set $\xi_i^1(\xi_i, m) \equiv m^* + v + (\hat{m}, \hat{\xi}_i)$, $\hat{m} \equiv m - m^1 e^1$ and $\hat{\xi}_i^\mu \equiv \xi_i^1 \xi_i^\mu$.

We now use the exponential Markov inequality and estimate the Laplace transform

$$I\!E \exp\left[t \mathbb{1}_{\{(\hat{\xi}_i, \hat{m}) < -\frac{7}{4} m^* - v\}} \left((\hat{\xi}_i, \hat{m}) + v\right)^2\right] \tag{3.38}$$

A rather straightforward computation with the choice $t = \frac{g}{2\|\hat{m}\|_2^2}$ gives

$$\begin{aligned}
&I\!E \exp\left[\frac{g}{2\|\hat{m}\|_2^2} \mathbb{1}_{\{(\hat{\xi}_i, \hat{m}) < -\frac{7}{4} m^* - v\}} \left((\hat{\xi}_i, \hat{m}) + v\right)^2\right] \\
&\leq 1 + \frac{1}{\sqrt{1-g}} \exp\left(-\frac{1}{2\|\hat{m}\|_2^2} \left[\left(\tfrac{7}{4} m^* + v\right)^2 - g\left(\tfrac{7}{4} m^*\right)^2\right]\right)
\end{aligned} \tag{3.39}$$

Since under our assumptions $\|\hat{m}\|_2^2 + v^2 \leq \frac{1}{4}(m^*)^2$ and $\|\hat{m}\|_2^2 + (m^* + v)^2 \geq (1 - \frac{1}{35})^2 (m^*)^2$, we have the bounds $v \geq -\left[\frac{1}{8} + \frac{1}{2}[(1 - \frac{1}{35})^2 - 1]\right] m^* \approx -0.096836 m^*$ and $\|\hat{m}\|_2^2 \leq (m^*)^2/4$, which gives with $g = \frac{1}{2}$,

$$I\!E \exp\left[\frac{1}{4\|\hat{m}\|_2^2} \mathbb{1}_{\{(\hat{\xi}_i, \hat{m}) < -\frac{7}{4} m^* - v\}} \left((\hat{\xi}_i, \hat{m}) + v\right)^2\right] \leq 1.10778 \tag{3.40}$$

Using that $\|m - e^1 m^*\|_2 \geq \|\hat{m}\|_2$, we get from here

$$\begin{aligned}
&I\!P\left[\frac{1}{N} \sum_i \mathbb{1}_{\{(\hat{\xi}_i, \hat{m}) < -\frac{3}{2} m^*/2 - v\}} \left((\hat{\xi}_i, \hat{m}) + v\right)^2 > y \|m - e^1 m^*\|_2^2\right] \\
&\leq \exp\left(-N \frac{y \|m - e^1 m^*\|_2^2}{4 \|\hat{m}\|_2^2} + N 0.1024\right) \\
&\leq \exp\left(-N \left[\frac{y}{4} - 0.1024\right]\right)
\end{aligned} \tag{3.41}$$

Choosing $y = 0.5$ then gives the assertion of Lemma 3.7. $\diamond$

We can now conclude the proof of Theorem 3.1. Note that statement (i) follows immediately from Corollary 3.4, if $c_1$ is sufficiently small to allow us to set $\epsilon(\alpha) = \frac{1}{35}$ and if $\bar{c}$ satisfies $\hat{c}(\beta) \geq \bar{c}(m^*)^2$.

Combining the estimates of Lemma 3.8 and 3.9, and choosing a constant $c_4 < 1/2$, we get that if only $c_1$ (and thus $r(\alpha)$) is sufficiently small,

$$I\!P\left[\Phi(m) - \phi(m^*) \geq \tilde{c}_2 (m^*)^4\right] \leq e^{-\tilde{c}_3 N} \tag{3.42}$$

for all $m \in D_{c_4 m^*, 1/35}$ and for some strictly positive constants $\tilde{c}_2$ and $\tilde{c}_3$. It remains only to extend this to an estimate of the supremum over all $m \in D_{c_4 m^*, 1/35}$. Let us choose $k > 2$. Then as in Section 2, we find get immediately that

$$I\!P\left[\sup_{m \in D_{c_4 m^*, 1/35} \cap W_{M, \alpha^k}} \Phi(m) - \phi(m^*) \geq \tilde{c}_2 (m^*)^4\right] \leq e^{N[\alpha(k|\ln \alpha|+1) - \tilde{c}_3]} \tag{3.43}$$

But by Lemma 3.5 the supremum over $D_{c_4 m^*, 1/35}$ differs from the lattice supremum by no more than $2\alpha^k$, so that the claim (ii) of the theorem follows by slightly adjusting the constants $\tilde{c}_2$ and $\tilde{c}_4$. $\diamond\diamond$

# 4. Local minima of $\Phi$ near the 'Mattis states'

We will now show that the large deviation function $\Phi(m)$ actually has a quadratic behaviour in the neighborhood of the minima that correspond to the stored patterns. We already know that for very small $\alpha$, the absolute minima of $\Phi$ are located in the vicinity of these points. Here we will compute the location of the minima more precisely, and we show that they exist for much larger values of $\alpha$ than those for which our proof in the previous section worked. The proofs in this section use some of the methods introduced in Section 2.

## 4.1. Upper and lower bounds on $\Phi$

Let us for convenience consider the minimum at $m^{(1,1)}$. We set

$$m = e^1 m^* + v \tag{4.1}$$

We recall the notation $A \equiv \xi^t \xi / N$ and $B \equiv A - \mathbb{1}$. We may write

$$\begin{aligned}
\Phi(m) &= -\tfrac{1}{2}(m, Bm) + \tfrac{1}{N} \sum_i \phi((\xi_i, m)) \\
&= -\tfrac{1}{2}(v, Bv) - (v, Bm^* e^1) + \tfrac{1}{N} \sum_i \phi\left(m^* \xi_i^1 + (\xi_i, v)\right) \\
&= -\tfrac{1}{2}(v, Bv) - m^*\left(v, z^{(1)}\right) + \tfrac{1}{N} \sum_i \phi\left(m^* + \xi_i^1 (\xi_i, v)\right)
\end{aligned} \tag{4.2}$$

where $z_1^{(1)} = 0$ and for $\mu \neq 1$, $z_\mu^{(1)} = \tfrac{1}{N} \sum_i \xi_i^\mu \xi_i^1$. Here $m^* \equiv m^*(\beta)$ is by assumption one of the values at which $\phi(x)$ attains its minimum. We have the following result on the function $\phi$

**Lemma 4.1:** *Assume that $|z| \leq \tau m^*$. Then, for all $\beta > 1$, there exists a constant $0 \leq c(\beta, \tau) < 1$ such that*

$$\phi(m^* + z) - \phi(m^*) \leq \frac{z^2}{2}\left[1 - \beta(1 - (m^*)^2)\right](1 + c(\beta, \tau)) \tag{4.3}$$

*and*

$$\phi(m^* + z) - \phi(m^*) \geq \frac{z^2}{2}\left[1 - \beta(1 - (m^*)^2)\right](1 - c(\beta, \tau)) \tag{4.4}$$

$c(\beta, \tau)$ *satisfies* $\lim_{\beta \downarrow 1} c(\beta, \tau) \leq \frac{\tau(1+\tau)}{2}$ *and* $\lim_{\beta \uparrow \infty} c(\beta, \tau) = 0$.

*Moreover, for all values of $z$,*

$$\phi(m^* + z) - \phi(m^*) \geq 0 \tag{4.5}$$

*and*

$$\phi(m^* + z) - \phi(m^*) \leq \frac{z^2}{2}\left[1 - \beta + \beta \tanh^2(\beta(m^* + |z|))\right] \tag{4.6}$$

**Proof:** Taylor's formula with remainder gives that

$$\phi(m^* + z) - \phi(m^*) - \frac{z^2}{2}\phi''(m^*) = \phi'''(\tilde{z})\frac{z^3}{6} \tag{4.7}$$

for some $\tilde{z} \in [m^*, m^* + z]$. Now

$$\phi''(\tilde{z}) = 1 - \beta(1 - \tanh^2(\beta\tilde{z})) \tag{4.8}$$

and

$$\phi'''(\tilde{z}) = 2\beta^2 \frac{\tanh(\beta\tilde{z})}{\cosh^2(\beta\tilde{z})} \tag{4.9}$$

Since $m^* = \tanh(\beta m^*)$ by definition, we get

$$\phi(m^* + z) - \phi(m^*) = \frac{z^2}{2}\left[1 - \beta(1 - (m^*)^2)\right]\left(1 + \frac{2z\beta^2 \frac{\tanh(\beta\tilde{z})}{\cosh^2(\beta\tilde{z})}}{6[1 - \beta(1 - (m^*)^2)]}\right) \tag{4.10}$$

For $\beta$ close to 1, a good estimate is

$$\left|\frac{2z\beta^2 \frac{\tanh(\beta\tilde{z})}{\cosh^2(\beta\tilde{z})}}{6[1 - \beta(1 - (m^*)^2)]}\right| \leq \frac{1}{3}\frac{\tau(1 + \tau)\beta^3 (m^*)^2}{1 - \beta + \beta (m^*)^2} \tag{4.11}$$

Since a simple calculation shows that to first order in $\beta - 1$, $(m^*(\beta))^2 = 3(\beta - 1)$, this gives the desired estimate in the case $\beta \downarrow 1$. For $\beta$ large, note first that $m^*(\beta) \uparrow 1$, exponentially fast, and so $(1 - \beta(1 - (m^*)^2) \sim 1 - \frac{e^{-\beta}}{\cosh\beta}$ tends to 1 exponentially fast. From this it is plain to see that in that case the right hand side of (4.11) is of the order of $\beta^2 / \cosh^2(\beta(m^*(1 - a))$ which tends to zero exponentially fast as $\beta \uparrow \infty$.

(4.5) is trivial and (4.6) follows from Taylor's theorem with second order remainder and (4.8).
◇

We would like to use the bounds from Lemma 4.1 in (4.2), and preferably the sharper bounds (4.3) and (4.4). The problem here is that even under smallness conditions on $v$ we cannot be sure that for all $i$ the quantities $(\xi_i, v)$ will have modulus smaller than $a \equiv \tau m^*$. We will first show how to deal with this for the lower bound. The proof of the upper bound will be similar but slightly more involved.

We get from Lemma 4.1 for $\Phi(m)$ the lower bound

$$\Phi(m) - \phi(m^*) \geq -\tfrac{1}{2}(v, Bv) - m^*\left(v, z^{(1)}\right)$$
$$+ \tfrac{1}{2N}\left[1 - \beta(1 - (m^*)^2)\right](1 - c(\beta,\tau))\sum_{i=1}^{N} \mathbb{I}_{\{|(\xi_i,v)|<a\}}(\xi_i, v)^2$$
$$= -\tfrac{1}{2}(v, Bv) - m^*\left(v, z^{(1)}\right)$$
$$+ \tfrac{1}{2N}\left[1 - \beta(1 - (m^*)^2)\right](1 - c(\beta,\tau))\sum_{i=1}^{N}(\xi_i, v)^2 \qquad (4.12)$$
$$- \tfrac{1}{2N}\left[1 - \beta(1 - (m^*)^2)\right](1 - c(\beta,\tau))\sum_{i=1}^{N} \mathbb{I}_{\{|(\xi_i,v)|\geq a\}}(\xi_i, v)^2$$
$$= \tfrac{1}{2}(v, [\mathbb{I} - (1 - c_-(\beta,\tau))A]v) - m^*\left(v, z^{(1)}\right)$$
$$- c_-(\beta)\tfrac{1}{2N}\sum_{i=1}^{N} \mathbb{I}_{\{|(\xi_i,v)|\geq a\}}(\xi_i, v)^2$$

where we have set $\left[1 - \beta(1 - (m^*)^2)\right](1 \pm c(\beta,\tau)) \equiv c_\pm(\beta,\tau)$ The last line in this bound is the only difficult one to treat. We set

$$X_a(v) \equiv \sum_{i=1}^{N}(\xi_i, v)^2 \mathbb{I}_{\{|(\xi_i,v)|\geq a\}} \qquad (4.13)$$

Our problem will be to estimate the supremum of this quantity over all $v$ in some ball. This problem is reminiscent to what we did in Section 2 when we estimated norms of the matrices $A$, and we will solve it in a very similar way. As we will see in the process of our analysis, we will also have to consider simultaneously the related variables

$$Y_a(v) \equiv \tfrac{1}{N}\sum_{i=1}^{N} \mathbb{I}_{\{|(\xi_i,v)|>a\}} \qquad (4.14)$$

As a starting point, we need estimates on the size of these random variables for fixed $v$. They are given by the following lemma.

**Lemma 4.2:** *Let $\xi_i^\mu$ be independent centered Bernoulli random variables. Define $p_a(x) \equiv 2\exp\left(-\tfrac{a^2}{2x^2}\right)$. Then*

$$\mathbb{P}\left[X_a(v) \geq xN\right] \leq \exp\left(N\left[2\sqrt{p_a(\|v\|_2)} - \tfrac{x}{4\|v\|_2^2}\right]\right) \qquad (4.15)$$

*and for $x \geq p_a(\|v\|_2)$*

$$\mathbb{P}\left[Y_a(v) \geq x\right] \leq \exp\left(-N\tfrac{(x - p_a(\|v\|_2))^2}{3p_a(\|v\|_2)}\right) \qquad (4.16)$$

**Proof:** We begin with the proof of (4.15). By the exponential Markov inequality we have that for any positive $t$,

$$IP\left[\sum_{i=1}^{N}(\xi_i,v)^2 1\!\!1_{\{|(\xi_i,v)|>a\}} \geq xN\right] \leq e^{-txN}\left[I\!\!E e^{t(\xi_i,v)^2 1\!\!1_{\{|(\xi_i,v)|>a\}}}\right]^N \qquad (4.17)$$

To estimate the Laplace transform, we write

$$I\!\!E e^{t(\xi_i,v)^2 1\!\!1_{\{|(\xi_i,v)|>a\}}} = I\!\!E 1\!\!1_{\{|(\xi_i,v)|\leq a\}} + I\!\!E e^{t(\xi_i,v)^2} 1\!\!1_{\{|(\xi_i,v)|>a\}}$$
$$\leq 1 + I\!\!E e^{t(\xi_i,v)^2} 1\!\!1_{\{|(\xi_i,v)|>a\}} \qquad (4.18)$$
$$\leq \exp\left(I\!\!E e^{t(\xi_i,v)^2} 1\!\!1_{\{|(\xi_i,v)|>a\}}\right)$$

For $t < 1/2\|v\|_2$ we have that

$$I\!\!E e^{t(\xi_i,v)^2} 1\!\!1_{\{|(\xi_i,v)|>a\}} \leq 2 I\!\!E e^{t(\xi_i,v)^2} 1\!\!1_{\{(\xi_i,v)>a\}}$$
$$\leq 2 \inf_{s\geq 0} e^{-sa} I\!\!E \int \frac{dz}{\sqrt{2\pi}} \exp\left(-\frac{z^2}{2} + [\sqrt{2t}z + s](\xi_i,v)\right)$$
$$= 2 \inf_{s\geq 0} \frac{\exp\left(-sa + \frac{s^2\|v\|_2^2}{2(1-2t\|v\|_2^2)}\right)}{\sqrt{1-2t\|v\|_2^2}} \qquad (4.19)$$
$$= \frac{2\exp\left(-\frac{a^2}{2\|v\|_2^2}(1-2t\|v\|_2^2)\right)}{\sqrt{1-2t\|v\|_2^2}}$$

Setting $t = \frac{1}{4\|v\|_2^2}$ we obtain (4.15).

To prove (4.16) we use again the exponential Markov inequality to get that for any positive $t$

$$IP[Y_a(v) \geq x] \leq e^{-txN}\left[I\!\!E e^{t1\!\!1_{\{|(\xi_1,v)|>a\}}}\right]^N$$
$$= e^{-txN}\left[(e^t - 1)IP(|(\xi_1,v)| > a) + 1\right]^N \qquad (4.20)$$

Now

$$IP((\xi_1,v) > a) \leq e^{-\frac{a^2}{2\|v\|_2^2}} \qquad (4.21)$$

Thus, since $(\xi_1,v)$ is a symmetric r.v. we get for $x \geq p_a(\|v\|_2)$

$$IP[Y_a(v) \geq x] \leq \inf_{t\geq 0} \exp\left\{-N\left[tx - \ln\left((e^t - 1)p_a(\|v\|_2) + 1\right)\right]\right\}$$
$$= \exp\left(-N I_{p_a(\|v\|_2)}(x)\right) \qquad (4.22)$$

where $I_p(x)$, for $p \in (0,1)$ is the well-known entropy function

$$I_p(x) \equiv \begin{cases} x\ln\left(\frac{x}{p}\right) + (1-x)\ln\left(\frac{1-x}{1-p}\right) & \text{, if } x \in [0,1] \\ \infty & \text{, if } x > 1 \end{cases} \qquad (4.23)$$

Finally, we use that (see [BG1])
$$I_p(x) \geq \frac{(x-p)^2}{3p} \tag{4.24}$$
to arrive at (4.16). ◊

Just as in Section 2 we can extract trivially bounds over lattice suprema. We get

**Lemma 4.3:** *Under the hypothesis of Lemma 4.2,*
$$\begin{aligned} &I\!\!P\left[\sup_{v\in\mathcal{W}_{M,r}(\rho)} X_a(v) \geq 4\rho^2\left[2\sqrt{p_a(\rho)} + \alpha(\ln(\rho/r)+c)\right]\right] \\ &\leq e^{-N(c-1)\alpha} \end{aligned} \tag{4.25}$$

*and*
$$I\!\!P\left[\sup_{v\in\mathcal{W}_{M,r}(\rho)} Y_a(v) \geq p_a(\rho) + \sqrt{3p_a(\rho)\alpha(\ln(\rho/r)+c)}\right] \leq \exp\{-N(c-1)\alpha\} \tag{4.26}$$

**Proof:** From Lemma 2.1 we have that
$$\begin{aligned} &I\!\!P\left[\sup_{v\in\mathcal{W}_{M,r}(\rho)} X_a(v) \geq xN\right] \\ &\leq e^{\alpha N(\ln_+\frac{\rho}{r}+1)}\sup_{v:\|v\|_2\leq\rho} I\!\!P\left[\sum_{i=1}^N (\xi_i,v)^2 \mathbb{I}_{\{|(\xi_i,v)|>a\}} \geq xN\right] \end{aligned} \tag{4.27}$$

We use Lemma 4.2 and choose $x$ sufficiently large that the resulting probability offsets the exponential prefactor. For this we set
$$x = 4\rho^2\left[2\sqrt{p_a(\rho)} + \alpha(\ln(\rho/r)+c)\right] \tag{4.28}$$
This gives (4.25) immediately. (4.26) follows in the same way. ◊

Now let $D \subset I\!\!R^M$ be any bounded domain. Our aim is to get estimates on quantities like $\sup_{v\in D} Y_a(v)$. As in Section 2 we note that $v \in D$ can be represented in the form
$$v = \sum_{n=1}^\infty v_n \tag{4.29}$$
with $v_n \in \mathcal{W}_{M,r_n}(r_{n-1}) \equiv \mathcal{W}(n)$ for $n > 1$ and $v_1 \in D \cap \mathcal{W}_{M,r_1}$.

The following observation is crucial:

**Lemma 4.4:** *Let $a_1 = a - d_1$, $d_1 \ll a$ be positive real constants. Then*
$$X_a(v_1+\epsilon) \leq X_{a_1}(v_1) + 2\sqrt{X_{a_1}(v_1)}\sqrt{(\epsilon,A\epsilon)} + 2a_1^2 Y_{d_1}(\epsilon) + 3(\epsilon,A\epsilon) \tag{4.30}$$

*and*

$$Y_a(v_1 + \epsilon) \leq Y_{a_1}(v_1) + Y_{d_1}(\epsilon) \tag{4.31}$$

**Proof:** The proof is based on the trivial observation that

$$\{|(\xi_i, (v_1 + \epsilon))| > a\}$$
$$= \{|(\xi_i, v_1)| \leq a_1\} \cap \{|(\xi_i, (v_1 + \epsilon))| > a\} \cup \{|(\xi_i, v_1)| > a_1\} \cap \{|(\xi_i, (v_1 + \epsilon))| > a\} \tag{4.32}$$
$$\subset \{|(\xi_i, v_1)| \leq a_1\} \cap \{|(\xi_i, \epsilon)| > d_1\} \cup \{|(\xi_i, v_1)| > a_1\}$$

This gives,

$$\begin{aligned}
\mathbb{I}_{\{|(\xi_i,(v_1+\epsilon))|>a\}}(\xi_i,(v_1+\epsilon))^2 &\leq \mathbb{I}_{\{|(\xi_i,v_1)|>a_1\}}((\xi_i,v_1)+(\xi_i,\epsilon))^2 \\
&\quad + \mathbb{I}_{\{|(\xi_i,v_1)|\leq a_1\}}\mathbb{I}_{\{|(\xi_i,\epsilon)|>d_1\}}((\xi_i,v_1)+(\xi_i,\epsilon))^2 \\
&\leq \mathbb{I}_{\{|(\xi_i,v_1)|>a_1\}}(\xi_i,v_1)^2 + (\xi_i,\epsilon)^2 + 2\mathbb{I}_{\{|(\xi_i,v_1)|>a_1\}}(\xi_i,v_1)(\xi_i,\epsilon) \\
&\quad + 2(\xi_i,\epsilon)^2 + 2a_1^2 \mathbb{I}_{\{|(\xi_i,\epsilon)|>d_1\}}
\end{aligned} \tag{4.33}$$

where some of the indicator functions have been dropped carelessly, and the inequality $(a+b)^2 \leq 2a^2 + 2b^2$ was used in the term that we anticipate as being small. Performing the summation over $i$ and using the Schwarz-inequality in the second term we arrive at (4.30). (4.31) is simpler and follows in the same way. $\diamondsuit$

**Corollary 4.5:** *Assume that $D \subset \mathbb{R}^M$ is sufficiently regular s.t. $D \subset \bigcup_{x \subset W_{M,r_1} \cap D} B_{r_1}(x)$, where $B_r(x)$ is the ball of radius $r$ centered at $x$ and set $B_r \equiv B_r(0)$. Then*

$$\sup_{v \in D} X_a(v) \leq \left(\sqrt{\sup_{v_1 \subset W_{M,r_1} \cap D} X_{a_1}(v_1)} + r_1\sqrt{\|A\|}\right)^2 + 2a_1^2 \sup_{\epsilon \in B_{r_1}} Y_{d_1}(\epsilon) + 2r_1^2 \|A\| \tag{4.34}$$

*and*

$$\sup_{v \in D} Y_a(v) \leq \sup_{v_1 \subset W_{M,r_1} \cap D} Y_{a_1}(v_1) + \sup_{\epsilon \in B_{r_1}} Y_{d_1}(\epsilon) \tag{4.35}$$

**Proof:** This is an immediate consequence of the previous considerations and the fact that $\sup_{\epsilon \in B_{r_1}} (\epsilon, A\epsilon) \leq r_1^2 \|A\|$ by the definition of the norm. $\diamondsuit$

Clearly, the representation of the supremum can serve as a starting point for an iteration. The norm of the matrix $A$ has been estimated in Section 2 and we know that it is close to one (for small $\alpha$) with probability exponentially close to one. The supremum over $W_{M,r_1}$ is a lattice supremum and has already been estimated. The remaining term is a supremum over a much smaller domain as before, and by repeated application of (4.35) will be shown to be very small. We formulate this in the next lemma.

**Lemma 4.6:**

$$I\!P\left[\sup_{v \in B_{r_0}} X_a(v) \geq r_0^2\left(2\sqrt{2\sqrt{p_{a_1}(r_0)} + \alpha(|\ln\frac{r_0}{r_1}| + c)} + \frac{r_1}{r_0}\sqrt{\|A\|}\right)^2 + 2r_1^2\|A\| + 2a_1^2 y\right]$$
$$\leq e^{-\alpha N(c-1)} + I\!P\left[\sup_{\epsilon \in B_{r_1}} Y_{d_1}(\epsilon) \geq y\right] \quad (4.36)$$

**Proof:** This is an immediate consequence of Corollary 4.5 and Lemma 4.4.◇

For the last term in the bound (4.36) we get the following

**Lemma 4.7:** *Suppose that* $\frac{r_1}{r_0} \ll d_1$

$$I\!P\left[\sup_{\epsilon \in B_{r_1}} Y_{d_1}(\epsilon) \geq e^{-\frac{1}{4}\left(\frac{d_1}{r_1}\right)^2(1-\sqrt{\alpha})^2}\left(2\left\{e^{-\frac{1}{4}\left(\frac{d_1}{r_1}\right)^2(1-\sqrt{\alpha})^2} + \sqrt{3\alpha(|\ln\alpha| + C)}\right\} + \frac{3}{\sqrt{N}}\right)\right]$$
$$\leq 2\exp\{-N(C-1)\alpha\} \quad (4.37)$$

**Proof:** By the same type of considerations as above, using in particular (4.35), we get that

$$\sup_{\epsilon \in B_{r_1}} Y_{d_1}(\epsilon) \leq \sum_{k=2}^{\infty} \sup_{\epsilon_k \subset \mathcal{W}(k)} Y_{b_k}(\epsilon_k) \quad (4.38)$$

where $b_k$ is some decreasing sequence of positive numbers that satisfies $\sum_{k=2}^{\infty} b_k = d_1$. Note that $|\mathcal{W}(k)| \leq e^{\alpha N(|\ln\frac{r_{k-1}}{r_k}|+1)}$, and so, using Lemma 4.2, we have

$$I\!P\left[\sup_{\epsilon_k \subset \mathcal{W}(k)} Y_{b_k}(\epsilon_k) \geq p_{b_k}(r_{k-1}) + \sqrt{3p_{b_k}(r_{k-1})\alpha\left(|\ln\frac{r_{k-1}}{r_k}| + \zeta_k\right)}\right] \leq e^{-N\alpha(\zeta_k - 1)} \quad (4.39)$$

At this point one can make some reasonable choice for the parameters $r_k$, $b_k$ and $\zeta_k$. We will set

$$r_k = \alpha^{k-1} r_1$$
$$b_k = \alpha^{(k-2)/2} d_1(1 - \sqrt{\alpha}) \quad (4.40)$$
$$\zeta_k = c + \frac{k^2}{\alpha N}$$

To simplify our expressions we will assume in the sequel that

$$\exp\left(-\frac{d_1^2}{4r_1^2}(1-\sqrt{\alpha})^2\right) \leq \frac{1}{2} \quad (4.41)$$

and that $\alpha < 1/2$. Then by a straightforward computation

$$\sum_{k=2}^{\infty}\left[p_{b_k}(r_{k-1}) + \sqrt{3p_{b_k}(r_{k-1})\alpha\left(|\ln\frac{r_{k-1}}{r_k}| + \zeta_k\right)}\right]$$
$$\leq e^{-\frac{1}{4}\left(\frac{d_1}{r_1}\right)^2(1-\sqrt{\alpha})^2}\left(2\left\{e^{-\frac{1}{4}\left(\frac{d_1}{r_1}\right)^2(1-\sqrt{\alpha})^2} + \sqrt{3\alpha(|\ln\alpha|+1)}\right\} + \frac{3}{\sqrt{N}}\right) \quad (4.42)$$

Form this, the Lemma follows immediately from the observation that the probability that a sum of r.v.'s exceeds a given sum only if at least one of the r.v.'s exceeds the corresponding summand. ◊

**Proposition 4.8:** *Define*

$$\Gamma(\alpha, a, \rho) = \left(2\sqrt{2\sqrt{2}e^{-\frac{(1-3\sqrt{\alpha})^2}{(1-\sqrt{\alpha})^2}\frac{a^2}{4\rho^2}} + \alpha(|\ln\alpha|+2) + \alpha\sqrt{1+r(\alpha)}}\right)^2 + 2\alpha^2(1+r(\alpha)) \tag{4.43}$$

$$+ \tfrac{1}{2}\alpha\left(2e^{-\frac{a^2}{\alpha\rho^2}} + 2\sqrt{3\alpha(|\ln\alpha|+2)}\right)$$

*Then*

$$I\!P\left[\sup_{v\in B_\rho} X_a(v) \geq \rho^2 \Gamma(\alpha, a, \rho)\right] \leq e^{-\alpha N} + I\!P[\|A - 1\!\!1\| \geq r(\alpha)] \tag{4.44}$$

**Proof:** The proposition is just a combination of Lemma 4.7 and 4.8 and a somewhat arbitrary choice of $r_1$ and $d_1$. If we set $a_1 = (1-\lambda)a$, $d_1 = \lambda a$ and $r_1 = \alpha\rho$. Then

$$I\!P\left[\sup_{v\in B_\rho} X_a(v) \geq \rho^2 \left(2\sqrt{2\sqrt{2}e^{-\frac{(1-\lambda)^2 a^2}{4\rho^2}} + \alpha(|\ln\alpha|+c) + \alpha\sqrt{\|A\|}}\right)^2 + 2\alpha^2\rho^2\|A\| \right.$$

$$\left. + 2(1-\lambda)^2 a^2 e^{-\frac{1}{4}\left(\frac{\lambda^2}{\alpha^2}\right)^2(1-\sqrt{\alpha})^2} \left(2e^{-\frac{\lambda^2}{4\alpha^2}\left(\frac{a}{\rho}\right)^2(1-\sqrt{\alpha})^2}\right) + 2\sqrt{3\alpha(|\ln\alpha|+c)}\right] \tag{4.45}$$

$$\leq 2e^{-N\alpha(c-1)}$$

Finally, we may choose $\lambda$ in such a way that $\frac{\lambda^2}{4\alpha}(1-\sqrt{\alpha})^2 = 1$ and this together with the estimate on the norm of $A$ from Section 2 gives the proposition. ◊

We combine the previous results to get the desired lower bound

$$\Phi(m^*e^1 + v) - \phi(m^*) \geq \tfrac{1}{2}(v, B_-(\rho)v) - m^*(v, z^{(1)}) \tag{4.46}$$

with

$$B_-(\rho) \equiv c_-(\beta,\tau)1\!\!1 + (1\!\!1 - A)(1 - c_-(\beta,\tau)) - c_-(\beta,\tau)\Gamma(a,\alpha,\rho)1\!\!1 \tag{4.47}$$

We turn now to the derivation of the corresponding upper bound. The strategy to use will depend on the value of $\beta$. If $\beta \geq 1.5$, then $m^*(\beta) \geq 0.5$ and little is lost if we use instead of (4.6) the rougher estimate

$$\phi(m^* + z) - \phi(m^*) \leq \frac{z^2}{2} \tag{4.48}$$

This then yields

$$\Phi(m) - \phi(m^*) \leq -\tfrac{1}{2}(v, Bv) - m^*\left(v, z^{(1)}\right)$$

$$+ \tfrac{1}{2N}\left[1 - \beta(1-(m^*)^2)\right](1 + c(\beta,\tau))\tfrac{1}{N}\sum_{i=1}^{N}\mathbb{1}_{\{|(\xi_i,v)|<a\}}(\xi_i,v)^2$$

$$+ \tfrac{1}{2N}\sum_{i=1}^{N}\mathbb{1}_{\{|(\xi_i,v)|\geq a\}}(\xi_i,v)^2 \tag{4.49}$$

$$= \tfrac{1}{2}\left(v, [\mathbb{1} - (1 - c_+(\beta,\tau))A]v\right) - m^*\left(v, z^{(1)}\right)$$

$$+ \tfrac{1}{2}X_a(v)\left[1 - c_+(\beta,\tau)\right]$$

From the previous estimates on $X_a(v)$ it then follows immediately that

$$\Phi(m^*e^1 + v) - \phi(m^*) \leq \tfrac{1}{2}(v, B_+(\rho)v) - m^*(v, z^{(1)}) \tag{4.50}$$

with

$$B_+(\rho) \equiv c_+(\beta,\tau)\mathbb{1} + (\mathbb{1} - A)(1 - c_+(\beta,\tau))$$
$$+ \mathbb{1}[1 - c_+(\beta,\tau)]\Gamma(a,\alpha,\rho) \tag{4.51}$$

For $\beta$ close to 1, this estimate is not very good. This can be seen from the fact that in the difference between $B_-$ and $B_+$ there occurs a term that is not proportional to $(m^*)^2$. To remedy the situation we must proceed more carefully with the term $\tanh^2 \beta(m^* + |z|)$ in (4.6), taking advantage, on the other hand, of the fact that $\beta$ is now strictly bounded.

Thus we replace (4.49) by

$$\Phi(m) - \phi(m^*) \leq -\tfrac{1}{2}(v, Bv) - m^*\left(v, z^{(1)}\right)$$

$$+ \tfrac{1}{2N}\left[1 - \beta(1-(m^*)^2)\right](1 + c(\beta,\tau))\tfrac{1}{N}\sum_{i=1}^{N}\mathbb{1}_{\{|(\xi_i,v)|<a\}}(\xi_i,v)^2$$

$$+ (1-\beta)\tfrac{1}{2N}\sum_{i=1}^{N}\mathbb{1}_{\{|(\xi_i,v)|\geq a\}}(\xi_i,v)^2 + \tfrac{\beta}{2}X_{m^*K}(v)$$

$$+ \tfrac{\beta}{2N}\sum_{i=1}^{N}\mathbb{1}_{\{|(\xi_i,v)|\geq a\}}(\xi_i,v)^2\sum_{k=0}^{K-1}\mathbb{1}_{\{m^*k\leq|(\xi_i,v)|<(k+1)m^*\}}\tanh^2(\beta(m^* + |(\xi_i,v)|))$$

$$\leq \tfrac{1}{2}(v, [\mathbb{1} - (1 - c_+(\beta,\tau))A]v) - m^*\left(v, z^{(1)}\right)$$

$$+ \tfrac{1}{2}(1 - \beta + \beta\tanh^2(\beta m^*))X_a(v)$$

$$+ \tfrac{\beta}{2}\sum_{k=1}^{K-1}\tanh^2(\beta(m^*(k+1)))X_{km^*}(v) + \tfrac{\beta}{2}X_{m^*K}(v) \tag{4.52}$$

With our previous bounds, we can replace the various $X(v)$ by the bounds from Proposition 4.8 on their suprema over $v$ with given norm . To simplify the resulting expressions, we will use that for $\alpha \leq 0.1$,
$$\Gamma(\alpha, km^*, \rho) \leq 8\sqrt{2} \exp\left(-(1 - 2\sqrt{\alpha})^2 \frac{(km^*)^2}{4\rho^2}\right) + \alpha \left[4|\ln \alpha| + 10\right] \quad (4.53)$$

Moreover, we bound $\tanh^2(\beta(m^*(k+1))) \leq \beta^2(m^*)^2(k+1)^2$. Thus we can bound

$$\sup_{v:\|v\|_2 \leq \rho} \sum_{k=1}^{K-1} \tanh^2(\beta(m^*(k+1))) X_{km^*}(v)$$

$$\leq \beta^2 (m^*)^2 \rho^2 \sum_{k=1}^{K-1} (k+1)^2 8\sqrt{2} \exp\left(-(1 - 2\sqrt{\alpha})^2 \frac{(km^*)^2}{4\rho^2}\right) + \alpha K \left[4|\ln \alpha| + 10\right] \rho^2 \quad (4.54)$$

$$\leq \gamma (m^*)^2 \rho^2 K \sqrt{\alpha} \left[4|\ln \alpha| + 10\right] + 208\sqrt{2}\rho^2 (m^*)^2 \exp\left(-(1 - 2\sqrt{\alpha})^2 \frac{(m^*)^2}{4\rho^2}\right)$$

where the numerical constant in the last bound was obtained under the hypothesis that $\alpha$ and $\rho$ are such that $\exp\left(-(1 - 2\sqrt{\alpha})^2 \frac{(m^*)^2}{4\rho^2}\right) \leq 1/2$. Finally, $K$ must be chosen such that both

$$K\sqrt{\alpha} \left[4|\ln \alpha| + 10\right] \leq 1 \quad (4.55)$$

and

$$8\sqrt{2} \exp\left(-(1 - 2\sqrt{\alpha})^2 \frac{(Km^*)^2}{4\rho^2}\right) \leq \gamma \rho^2 (m^*)^2 \quad (4.56)$$

It is easy to check that this is the case if

$$K = \frac{2\rho}{m^*(1 - 2\sqrt{\alpha})} \sqrt{\left|\ln \frac{\gamma(\rho m^*)^2}{8\sqrt{2}}\right|} \quad (4.57)$$

Combining everything, we see that we get again the upper bound (4.50), but this time with

$$B_+(\rho) \equiv c_+(\beta, \tau) \mathbb{I} + (\mathbb{I} - A)(1 - c_+(\beta, \tau))$$
$$+ (m^*)^2 \beta \mathbb{I} \left[ \left((1 - \beta)/(m^*)^2 + \beta^2\right) \Gamma(\alpha, a, \rho) + 2\gamma + 300 \exp\left(-(1 - 2\sqrt{\alpha})^2 \frac{(m^*)^2}{4\rho^2}\right) \right] \quad (4.58)$$

We summarize the results of this subsection in the following theorem.

**Theorem 4.9:** *There exists a set $\tilde{\Omega} \subset \Omega$ of measure one such that for all but a finite number of values $N$, for any $0 < \rho < 1$ and for all $\|v\|_2 \leq \rho$,*

$$\Phi_{N,\beta}[\omega](m^* e^1 + v) - \phi(m^*) \geq \tfrac{1}{2}(v, B_-(\rho)v) - m^*(v, z^{(1)}) \quad (4.59)$$

*and*

$$\Phi_{N,\beta}[\omega](m^* e^1 + v) - \phi(m^*) \leq \tfrac{1}{2}(v, B_+(\rho)v) - m^*(v, z^{(1)}) \quad (4.60)$$

*where $B_-(\rho)$ is defined by (4.47) and $B_+(\rho)$ is given by (4.51) if $\beta > 1.1$ and by (4.58) if $1 < \beta \leq 1.1$.*

**Proof:** This theorem follows simply from our previous estimates and using the Borel-Cantelli lemma. ◊

### 4.2. Localization of the minima

Theorem 4.9 contains the main information needed for the analysis of the structure of the minima of the function $\Phi$. As we will explain later, it also serves as a starting point for a more refined analysis of that function.

**Theorem 4.10:** *There exist finite positive constants $c_1, c_2, c_3$ such that the following holds for almost all $\omega$ for all but a finite number of values $N$: If $\sqrt{\alpha} \leq c_1 (m^*(\beta))^2$ then for all $v$ such that*

$$c_2 \frac{\sqrt{\alpha} m^*(\beta)}{c_-(\beta, \tau)} \leq \|v\|_2 \leq c_3 m^*(\beta) \tag{4.61}$$

*and for all $(\mu, s)$,*

$$\Phi_{N,\beta}[\omega](sm^* e^\mu + v) > \phi(sm^* e^\mu) \tag{4.62}$$

**Remark:** Theorem 4.10 establishes the existence of a local minimum at a distance of order $\frac{\sqrt{\alpha} m^*(\beta)}{c_-(\beta,\tau)}$ from the points $sm^* e^\mu$. We will soon localize them more precisely. This is a generalization of the results of Newman [N] and Komlos and Paturi [KPa] to finite temperatures. If we consider the asymptotic regime where $\beta \sim \beta_c = 1$ we have that $m^*(\beta) \sim \sqrt{\beta - 1}$ and $c_-(\beta, \tau) \sim \beta^2 - 1$. The condition on $\alpha$ is then of the form $\alpha \leq c(\beta - 1)^2$ and for sufficiently small $c_1$ the upper bound is seen to be a multiple of the lower one. Notice that this behaviour of the critical $\alpha$ as a function of $\beta$ near one is (up to the constants) the same as the one found by [AGS] using the replica method. For large $\beta$, we have checked numerically that the constant $c_1$ can be chosen at least as 0.04.

**Proof:** The proof of Theorem 4.10 relies on the lower bound (4.59) from Theorem 4.9 and the following estimate on the norm of the vectors $z^{(\mu)}$.

**Lemma 4.11:** *Let $z_\mu^{(\nu)} \equiv \frac{1}{N} \sum_i \xi_i^\mu \xi_i^\nu$, for $\mu \neq \nu$ and $z_\nu^{(\nu)} \equiv 0$. Then, for all $\epsilon$ sufficiently small*

$$I\!P \left[ \left\| z^{(\nu)} \right\|_2 \geq (1 + \epsilon) \sqrt{\alpha} \right] \leq e^{-\epsilon^2 M/5} \tag{4.63}$$

**Proof:** Note that for fixed $\nu$, $z_\mu^{(\nu)}$ are independent for different $\mu$. Moreover, for $\mu \neq \nu$ and the assumption of the Lemma, they are stochastically dominated by independent normal distributed random variables $z_\mu$. The bound (4.63) then follows by a simple application of the exponential Markov inequality. ◊

To prove the proposition, we may now choose $a$ in Proposition 4.8 in a suitable way. A possible choice is $a = m^*/3$. With this choice $c_-(\beta, \tau) \geq \frac{4}{9}\left(1 - \beta(1 - (m^*)^2)\right)$. For $\|v\|_2$ satisfying the upper bound in (4.61), we can make the terms $e^{-\frac{a^2}{K\rho^2}}$ appearing in $\Gamma(a, \alpha, \rho)$ as small as desired by choosing $c_3$ small, while the terms $\alpha(\ln \rho/r + 2)$ can also be made small by choosing $c_1$ small, and also all the terms of order $\sqrt{\alpha}$ are small compared to $c_-(\beta, \tau)$ under the assumption on $\alpha$. Thus we get effectively a bound

$$\Phi(se^\mu + v) - \phi(m^*) \geq \tfrac{1}{4}c_-(\beta, \tau)\|v\|_2^2 - \sqrt{\alpha}(1 + \epsilon)\|v\|_2 \tag{4.64}$$

and this is strictly positive if $v$ satisfies the lower bound in (4.61). Moreover, the lower bound in (4.61) is smaller than the upper one if $c_1$ is sufficiently small, so that our statement is not void. ◇◇

Theorem 4.10 will sharpened in the sense that we can locate more precisely the position of the true local minima.

**Lemma 4.12:** *For all $\rho$ sufficiently small such that $B_-(\rho)$ is strictly positive we define $v^{(\mu)} \equiv m^* B_+(\rho)^{-1} z^{(\mu)}$. Then for all $\mu$ and for all $v$ such that*

$$\|v - v^{(\mu)}\|_2 > m^*\sqrt{\alpha}(1+\epsilon)\|B_+^{-1}\| \left(\sqrt{\|B_-(\rho)^{-1}\| \|B_+(\rho) - B_-(\rho)\|} + 2\|B_-(\rho)^{-1}\| \|B_+(\rho) - B_-(\rho)\|\right) \tag{4.65}$$

*and $\|v\|_2 \leq \rho$ one has that, for almost all $\omega$, for all but a finite number of indices $N$,*

$$\Phi_{N,\beta}[\omega](e^\mu m^* + v) > \Phi_{N,\beta}[\omega](e^\mu m^* + v^{(\mu)}) \tag{4.66}$$

**Proof:** This lemma follows from Theorem 4.9 by some elementary algebra and Lemma 4.11. ◇

It remains to estimate the various norms appearing in (4.65). This is an elementary, but somewhat painful, exercise and we will just consider the two asymptotic regimes $\beta \downarrow 1$ and $\beta \uparrow \infty$. We collect these bounds, which are easily obtained from our previous estimates without going into the details of the proofs. We also, for sake of clarity, take the liberty to throw away all insignificantly small corrections.

**Lemma 4.13:** *Let us put $\sqrt{\alpha} = \gamma(m^*)^2$. Then we have for $1 > \tau \geq 0$ to be chosen later:*

*(i) To leading order in the limit $\beta \downarrow 1$*

$$\|B_+^{-1}\| \leq \|B_-^{-1}\| \leq \frac{3}{2}\frac{1}{(m^*)^2}\frac{1}{1 - \tau(\tau+1)/2 - 3\gamma - (1 - \tau(\tau+1)/2)\Gamma(\tau m^*, \alpha, \rho)} \tag{4.67}$$

*and*

$$\|B_+ - B_-\| \leq (m^*)^2 \left(\frac{\tau(\tau+1)}{3} + 4\Gamma(\alpha, \tau m^*, \rho) + 2\gamma + 300\exp\left(-\frac{(m^*)^2}{4\rho^2}\right)\right) \tag{4.68}$$

(ii) If $\beta \uparrow \infty$,

$$\|B_+^{-1}\| \leq \|B_-^{-1}\| \leq \frac{1}{1 - \Gamma(\tau, \alpha, \rho)} \qquad (4.69)$$

and

$$\|B_+ - B_-\| \leq \Gamma(\tau, \alpha, \rho) \qquad (4.70)$$

and to leading order in $\alpha$, for $\rho = \sqrt{\alpha}$,

$$\Gamma(\tau, \alpha, \rho) \approx 8\sqrt{2} \exp\left(-\frac{\tau^2}{4\alpha^2}(1 - 2\sqrt{\alpha})^2\right) + \alpha[|\ln \alpha| + 2] \qquad (4.71)$$

Note that by Theorem 4.10 we only need to consider the ball of radius $\rho = c_2 \frac{m^*\sqrt{\alpha}}{c_-(\beta,\tau)}$. In case (ii) we can choose $\tau$ arbitrary close to one to get the result

$$\|B_+ - B_-\|\|B_-^{-1}\| \sim 8\sqrt{2} \exp\left(-\frac{1}{4\alpha}(1 - 2\sqrt{\alpha})\right)^2 + 2\alpha[|\ln \alpha| + 2] \qquad (4.72)$$

In case (i) we still have to make our choice for the parameter $\tau$. Note that in that case $c_-(\beta, \tau) \approx \frac{2}{3}(m^*)^2(1 - \tau(\tau+1)/2)$ and, anticipating that we will chose $\tau$ small, we have $\rho \approx \frac{2c_2}{3}\gamma m^*$. Inserting this value in the bounds (4.67) and (4.68), we get

$$\|B_+ - B_-\|\|B_-^{-1}\| \sim \frac{3}{2}\left[\tau(\tau+1) + 32\sqrt{2} \exp\left(-\frac{9\tau^2}{16c_2^2\gamma^2}\right) + 2\gamma + 300 \exp\left(-\frac{9}{16c_2^2\gamma^2}\right)\right] \qquad (4.73)$$

With the natural choice $\tau^2 = \frac{32c_2^2}{9}\gamma^2|\ln\gamma|$ this gives

$$\|B_+ - B_-\|\|B_-^{-1}\| \sim \frac{\sqrt{32}}{3}c_2\gamma\sqrt{|\ln\gamma|} + 2\gamma + O(\gamma^2) \qquad (4.74)$$

If we notice further that the dominant part of the matrix $B_+$ is a multiple of the identity, we arrive at the following

**Theorem 4.14:** *For any $\beta > 1$ set $b(\beta) \equiv \frac{m^*(\beta)}{1 - \beta(1 - (m^*)^2)}$. There exists $\gamma_0 > 0$ such that for all $\gamma \leq \gamma_0$, for almost all $\omega$, for all but a finite number of indices $N$, the following holds: for all $v$ such that $\|v\|_2 \leq c_3 m^*$ and*

$$\left\|v - b(\beta)z^{(\mu)}\right\|_2 \geq c_4 m^* \gamma^{3/2} \sqrt{|\ln\gamma|} \qquad (4.75)$$

$$\Phi_{N,\beta}[\omega](m^* e^\mu + v) > \inf_{\|v\|_2 \leq c_3 m^*} \Phi_{N,\beta}[\omega](m^* e^\mu + v) \qquad (4.76)$$

*for some strictly positive constant $c_4$ and where $c_3$ is the same constant as in Theorem 4.10.*

This theorem allows us to locate quite precisely the (random) position of the lowest minimum of the function $\Phi$ in vicinity of any of the points $m^* e^\mu$. It is of interest to observe that in smaller regions these minima are even unique, i.e. there are no other local minima in the immediate vicinity of the 'Mattis states'. This is the main content of the last theorem of this section.

**Theorem 4.15:** Assume that $1 < \beta < \infty$. Then there exists $\alpha_0(\beta)$ and $\rho(\beta)$ such that if $\alpha \leq \alpha_0(\beta)$, with probability one for all but a finite number of indices $N$, $\Phi_{N,\beta}[\omega](m^*e^1 + v)$ is a twice differentiable and convex function of $v$ for all $v$ with $\|v\|_2 \leq \rho(\beta)$.

**Proof:** The differentiability for fixed $N$ is no problem. The non-trivial assertion of the theorem is the local convexity. We have that

$$D^2\Phi(m^*e^1 + v) = \mathbb{1} - A + \frac{1}{N}\sum_{i=1}^N \phi''(m^*\xi_i^1 + (\xi_i, v))\xi_i^t\xi_i$$

$$= \mathbb{1} - A + \frac{1}{N}\sum_{i=1}^N \mathbb{1}_{\{|(\xi_i,v)|\leq \tau m^*\}}\phi''(m^*\xi_i^1 + (\xi_i, v))\xi_i^t\xi_i \qquad (4.77)$$

$$+ \frac{1}{N}\sum_{i=1}^N \mathbb{1}_{\{|(\xi_i,v)|> \tau m^*\}}\phi''(m^*\xi_i^1 + (\xi_i, v))\xi_i^t\xi_i$$

The point here is that

$$\phi''(x) = 1 - \beta\left(1 - \tanh^2(\beta x)\right) \qquad (4.78)$$

so that $\phi''(x) \geq c$ if $|x| \geq \frac{1}{\beta}\tanh^{-1}\left(\sqrt{1 - \frac{1-c}{\beta}}\right)$. Moreover, for arbitrary $x$ we have that $\phi''(x) \geq 1 - \beta$. Thus if we set $\tau = \frac{1}{m^*\beta}\tanh^{-1}\left(\sqrt{\frac{1-c}{\beta}}\right) - 1$, denoting by $\lambda_{min}\left(D^2\phi(m^*e^1 + v)\right)$ the smallest eigenvalue of $D^2\phi(m^*e^1 + v)$, we get that

$$\lambda_{min}\left(D^2\phi(m^*e^1 + v)\right) \geq 1 - (1-c)\|A\| - (\beta - 1 - c)\left\|\frac{1}{N}\sum_{i=1}^N \mathbb{1}_{\{|(\xi_i,v)|>\tau m^*\}}\xi_i^t\xi_i\right\| \qquad (4.79)$$

Denoting by $\lambda_{min}\left(D^2\phi(m^*e^1 + v)\right)$ the smallest eigenvalue of $D^2\phi(m^*e^1 + v)$ What we need to do is to estimate the norm of the last term in (4.79). Now,

$$\sup_{v\in B_\rho}\left\|\frac{1}{N}\sum_{i=1}^N \mathbb{1}_{\{|(\xi_i,v)|>\tau m^*\}}\xi_i^t\xi_i\right\| = \sup_{v\in B_\rho}\sup_{w\in S_\rho}\frac{1}{\rho^2}\frac{1}{N}\sum_{i=1}^N \mathbb{1}_{\{|(\xi_i,v)|>\tau m^*\}}(\xi_i, w)^2$$

$$\leq \frac{1}{\rho^2}\sup_{v\in B_\rho}\sup_{w\in B_\rho}\frac{1}{N}\sum_{i=1}^N \mathbb{1}_{\{|(\xi_i,v)|>\tau m^*\}}(\xi_i, w)^2 \qquad (4.80)$$

Using the trick to write

$$\mathbb{1}_{\{|(\xi_i,v)|>\tau m^*\}}(\xi_i, w)^2 = \mathbb{1}_{\{|(\xi_i,v)|>\tau m^*\}}(\xi_i, w)^2\left(\mathbb{1}_{\{|(\xi_i,w)|<|(\xi_i,v)|\}} + \mathbb{1}_{\{|(\xi_i,w)|\geq|(\xi_i,v)|\}}\right)$$

$$\leq \mathbb{1}_{\{|(\xi_i,v)|>\tau m^*\}}(\xi_i, v)^2 + \mathbb{1}_{\{|(\xi_i,w)|>\tau m^*\}}(\xi_i, w)^2 \qquad (4.81)$$

so that

$$\frac{1}{N}\sum_{i=1}^N \mathbb{1}_{\{|(\xi_i,v)|>\tau m^*\}}(\xi_i, w)^2 = X_{\tau m^*}(v) + X_{\tau m^*}(w) \qquad (4.82)$$

by which token we are reduced to estimate the same quantities as before. We obtain therefore on $\Omega_1$ for all $v$ with norm less than $\rho$,

$$\lambda_{min}\left(D^2\phi(m^*e^1+v)\right) \geq 1-(1-c)(1+r(\alpha))-(\beta-1-c)\Gamma(\alpha,\tau m^*,\rho) \qquad (4.83)$$

which proves the theorem and allows to estimate the constants involved.$\Diamond$

**Remark:** Note that the estimates derived from (4.83) become quite bad if $\beta$ is large. This is due to the fact that the second derivative of $\phi$ satisfies a poor uniform bound in this case. However, this bad bound is realized only in a small region, so that a more careful analysis should allow to replace $(1-\beta)$ by a bounded constant.

### 4.3. The macroscopic component of the minima near the 'Mattis states'

We have seen so far that the location of the minima of $\Phi$ is shifted away from the 'Mattis states' $m^*e^\mu$ by a random vector $z^{(\mu)}$, up to error terms of small norm. The components of $z^{(\mu)}$ are all "microscopic" i.e. of order $\frac{1}{\sqrt{N}}$. [AGS] found, on the basis of the replica method that the location of the minimum associated to the pattern $\mu$ undergoes a macroscopic shift of order $\exp\left(-\frac{1}{2\alpha}\right)$ of its $\mu$-th component. We will show that from Theorem 4.14 such a result can be derived in a rigorous form. Without restriction of generality, we consider a minimum with $(\mu=1,s=+1)$. We denote the 1-component of the location of a minimum according to Theorem 4.14 by $m^1(N)$ and set

$$m_+^1 \equiv \limsup_{N\uparrow\infty} m^1(N) \qquad (4.84)$$

and

$$m_-^1 \equiv \liminf_{N\uparrow\infty} m^1(N) \qquad (4.85)$$

**Theorem 4.16:** *Assume that $\alpha$ satisfies the hypothesis of Theorem 4.14. Then there exists a finite constant $c_5$ such that, $I\!\!P$-almost certainly,*

$$m_+^1 \leq \frac{1}{\sqrt{2\pi}}\int e^{-\frac{x^2}{2}}\tanh\beta(m_+^1+\sqrt{\gamma}m^*+\sqrt{\alpha}m^*x)dx + c_5\sqrt{\alpha}|\ln\alpha|e^{-\frac{1}{4\gamma^2|\ln\gamma|}} \qquad (4.86)$$

*and*

$$m_-^1 \geq \frac{1}{\sqrt{2\pi}}\int e^{-\frac{x^2}{2}}\tanh\beta(m_-^1+\sqrt{\gamma}m^*+\sqrt{\alpha}m^*x)dx - c_5\sqrt{\alpha}|\ln\alpha|e^{-\frac{1}{4\gamma^2|\ln\gamma|}} \qquad (4.87)$$

A special case of this Theorem is the following

**Corollary 4.17:** *In the limit $\beta \uparrow \infty$, $I\!P$-almost surely*

$$m^1_+ \leq Erf\left(\frac{m^1_+ + \sqrt{\alpha}}{\sqrt{2\alpha}}\right) + c_5\sqrt{\alpha}|\ln\alpha|e^{-\frac{1}{4\alpha^2|\ln\alpha|}} \tag{4.88}$$

*and*

$$m^1_- \geq Erf\left(\frac{m^1_- - \sqrt{\alpha}}{\sqrt{2\alpha}}\right) - c_5\sqrt{\alpha}|\ln\alpha|e^{-\frac{1}{4\alpha^2|\ln\alpha|}} \tag{4.89}$$

*where* $Erf(x) \equiv \int_0^x dt\, e^{-t^2}$ *is the error function.*

**Remark:** The bounds (4.89) can be evaluated numerically, but it is clear that (4.88) implies that $m^1_+ \leq 1 - O\left(e^{-1/\alpha}\right)$ and that for $\alpha$ small enough there exist $m^1_-$ of the same order which verifies (4.89). A numerical analysis of these inequalities shows that solutions near 1 exist up to values of $\alpha$ of order 0.1, much larger than those for which the hypothesis of Theorem 4.16 can be proven. Corollary (4.17) should be compared to the heuristically derived set of equations (4.5-7) of [AGS], namely $m^1 = Erf(m^1/\sqrt{2\alpha r})$, where $r = (1-C)^{-2}$ and $C = \sqrt{2/\pi\alpha r}\exp(-(m^1)^2/2\alpha r)$. They use these to determine the critical storage capacity by finding the maximal value $\alpha$ for which a non-zero solution exists. The inequalities of Theorem 4.16 compare with the equations (5.5,6) of [AGS].

**Proof:** Let $m$ be any minimum of $\Phi$ in the ball $B_{c_4\gamma b(\beta)\sqrt{\alpha}}(m^*(\beta)z^{(1)})$. Then, it must be a solution of the system of equations

$$m^\mu = \frac{1}{N}\sum_{i=1}^N \xi_i^\mu \tanh[\beta(\xi_i, m)] \quad , \quad \mu = 1, \ldots, M. \tag{4.90}$$

Now we can write $m = e^1 m^1 + m^* z^{(1)} + w$ where $w_1 \equiv 0$ and $\|w\|_2 \leq c_4\zeta b(\beta)\sqrt{\alpha}$. Then the equation for the component $m^1$ reeds

$$m^1 = \frac{1}{N}\sum_{i=1}^N \tanh[\beta(m^1 + m^*(\hat{\xi}_i, z^{(1)}) + (\hat{\xi}_i, w))] \tag{4.91}$$

where $\hat{\xi}_i \equiv \xi_i^1 \xi_i$. For any $a \geq 0$ we can write

$$m^1 \leq \frac{1}{N}\sum_{i=1}^N \tanh[\beta(m^1 + m^*(\hat{\xi}_i, z^{(1)}) + a)]$$

$$+ \frac{1}{N}\sum_{i=1}^N \mathbb{I}_{\{|(\hat{\xi}_i, w)| > a\}}\left[\tanh[\beta(m^1 + m^*(\hat{\xi}_i, z^{(1)}) + (\hat{\xi}_i, w))] - \tanh[\beta(m^1 + m^*(\hat{\xi}_i, z^{(1)}) + a)]\right]$$

$$\leq \frac{1}{N}\sum_{i=1}^N \tanh[\beta(m^1 + m^*(\hat{\xi}_i, z^{(1)}) + a)] + Y_a(w)$$

$$\tag{4.92}$$

where $Y_a(w)$ is defined in (4.14). Similarly

$$m^1 \geq \frac{1}{N} \sum_{i=1}^N \tanh[\beta(m^1 + m^*(\hat{\xi}_i, z^{(1)}) - a)] - Y_a(w) \tag{4.93}$$

We should expect that the averages over $i$ in the formulas (4.92) and (4.93) converge to expectations with respect to some measure. This is indeed the case due to the following lemma.

**Lemma 4.18:** *Let $\delta_x$ denote the Dirac-measure concentrated on $x$ and let $\mathcal{N}_{0,\alpha}$ be the centered Gaussian measure with mean zero and variance $\alpha$. Then,*

$$w - \lim_{N \uparrow \infty} \frac{1}{N} \sum_{i=1}^N \delta_{(\xi_i, z^{(1)})} = \mathcal{N}_{0,\alpha} \quad \mathbb{P}\text{-a.s.} \tag{4.94}$$

We will give the proof of this Lemma in the appendix.

We recall further that the quantity $\sup_{w \in B_\rho} Y_a(w)$ is known from Lemma 4.7 and, by a simple application of the Borel-Cantelli Lemma we obtain that, almost certainly,

$$\lim_{N \to \infty} \sup_{w \in B_{c_4 m^* \gamma^{3/2} \sqrt{|\ln \gamma|}}} Y_a(w) \leq \widetilde{\Gamma}(a, \alpha, m^*, \gamma) \tag{4.95}$$

where

$$\begin{aligned}\widetilde{\Gamma}(a, \alpha, m^*, \gamma) = \exp\left\{-\frac{1}{4}\left(\frac{a(1-\sqrt{\alpha})}{c_4 m^* \gamma^{3/2} \sqrt{|\ln \gamma|}}\right)^2\right\} \times \\ \left(2\left\{\exp\left\{-\frac{1}{4}\left(\frac{a(1-\sqrt{\alpha})}{c_4 m^* \gamma^{3/2} \sqrt{|\ln \gamma|}}\right)^2\right\} + \sqrt{3\alpha(|\ln \alpha| + C)}\right\}\right)\end{aligned} \tag{4.96}$$

Putting these observations together, we find that, almost surely,

$$m^1_+ \leq \frac{1}{\sqrt{2\pi\alpha}} \int dx\, e^{-\frac{x^2}{2\alpha}} \tanh[\beta(m^1_+ + a + m^* x)] + \widetilde{\Gamma}(a, \alpha, m^*, \gamma) \tag{4.97}$$

and

$$m^1_- \geq \frac{1}{\sqrt{2\pi\alpha}} \int dx\, e^{-\frac{x^2}{2\alpha}} \tanh[\beta(m^1_- - a + m^* x)] - \widetilde{\Gamma}(a, \alpha, m^*, \gamma) \tag{4.98}$$

Choosing $a = \sqrt{\gamma} m^*$ we obtain from here the claims of the theorem. ◊

## 5. Applications to the Gibbs measures: Proof of Theorem 3

Theorem 3 follows from the estimates in the last two sections in a fairly straightforward way along the lines of [BGP1] and [BGP2]. We only give a rough outline in order to avoid repetitions. In particular, we will only show how the results are obtained for the measures $\widetilde{\mathcal{Q}}$ and leave the remaining step that can be copied from [BGP1] to the reader. To simplify our notation, let us set $B_\rho^{(\mu)} \equiv B_\rho(m^* e^\mu)$ and $R_\rho \equiv \{\cup_{(\mu,s)} B_\rho(se^\mu m^*)\}^c$. Let us also introduce the integrals

$$I_\rho^{(\mu)} \equiv \int_{B_\rho^{(\mu)}} d^M z \, e^{-\beta N(\Phi(z)-\phi(m^*))} \tag{5.1}$$

and

$$J_\rho \equiv \int_{R_\rho} d^M z \, e^{-\beta N(\Phi(z)-\phi(m^*))} \tag{5.2}$$

Note that by symmetry, changing $B_\rho^{(\mu)}$ to $B_\rho(-m^* e^\mu)$ in (5.1) does not change $I_\rho^{(\mu)}$. To simplify our presentation, we will denote by $\Omega_2$ the subset of $\Omega$ on which our various bounds on $\Phi(m)$ from Sections 4 and 5 hold. All bounds stated in this section are true on $\Omega_2$; recall that the probability of $\Omega_2$ is exponentially close to one.

By Theorem 4.9 we have that

$$I_\rho^{(\mu)} \geq \int_{\|v\|_2 \leq \rho} d^M v \, e^{-\beta N \left(\frac{1}{2}(v, B_+(\rho)v) - m^*(v, z^{(\mu)})\right)} \tag{5.3}$$

Using that for $\rho \geq 4m^* \|z^{(\mu)}\|_2 \|B_+^{-1}(\rho)\|$

$$\int_{\|v\|_2 > \rho} d^M v \, e^{-\beta N \left(\frac{1}{2}(v, B_+(\rho)v) - m^*(v, z^{(\mu)})\right)} \leq \left[\frac{8\pi \|B_+^{-1}(\rho)\|}{\beta N}\right]^{M/2} \exp\left(-\frac{\beta N}{8\|B_+^{-1}(\rho)\|}\rho^2\right) \tag{5.4}$$

we get

$$I_\rho^{(\mu)} \geq \left[\frac{2\pi \|B_+^{-1}(\rho)\|}{\beta N}\right]^{M/2} \left(1 - 2^M e^{-\frac{\beta N}{8\|B_+^{-1}(\rho)\|}\rho^2}\right) \tag{5.5}$$

Using in addition to Theorem 4.9 the lower bounds on $\Phi$ from Section 4 we get on the other hand

$$\begin{aligned} J_\rho &\leq \int_{\Gamma_{1/35}} d^M z \, \exp\left(-\tfrac{1}{2}\beta N \bar{c}(m^*)^2 (\|z\|_2 - m^*)^2\right) \\ &\quad + \int_{D_{c_4 m^*, 1/35}} d^M z \, \exp\left(-\beta N c_2 (m^*)^4\right) \\ &\quad + 2 \sum_\mu \int_{\|v\|_2 \geq \rho} d^M v \, e^{-\beta N \left(\frac{1}{2}(v, B_- v) - m^*(v, z^{(\mu)})\right)} \\ &\leq (m^*)^M V_M \exp\left(-N\beta \tilde{c}(m^*)^4\right) + 2M \left[\frac{8\pi \|B_-^{-1}(\rho)\|}{\beta N}\right]^{M/2} e^{-\frac{\beta N}{8\|B_-^{-1}(\rho)\|}\rho^2} \end{aligned} \tag{5.6}$$

where $V_M \equiv \frac{2\pi^{M/2}}{\Gamma(M/2)}$ denotes the surface area of the $M$-dimensional unit sphere and for some constant $\tilde{c} > 0$. Now choose $\rho = c_5\sqrt{\alpha}/m^* = c_5\gamma m^*$. Set further $\|B_\pm^{-1}(\rho)\| = \tilde{c}_\pm/(m^*)^2$. Then the above estimates combine to

$$\frac{J_\rho}{I_\rho^{(\mu)}} \leq \beta^{M/2}\gamma^{-M}e^{M/2}\tilde{c}_+^{M/2}e^{-\frac{\beta M}{2\gamma^2}\tilde{c}} + 2M\left(\frac{4\tilde{c}_-}{\tilde{c}_+}\right)^{M/2}e^{-\frac{\beta N\gamma^2 c_5^2(m^*)^4}{8\tilde{c}_-}} \qquad (5.7)$$
$$\leq e^{-\beta M c_7}$$

where $c_7 > 0$ is some constant depending on $c_5$, $\gamma_a$ and $\tilde{c}_+/\tilde{c}_-$. Since clearly

$$\widetilde{\mathcal{Q}}(R_\rho) \leq \frac{J_\rho}{I_\rho^{(1)}} \qquad (5.8)$$

$\widetilde{\mathcal{Q}}(R_{c_5\gamma m^*})$ obeys the same bound.

Next, to prove the second statement of Theorem 3, observe that

$$\ln \frac{I_\rho^{(\mu)}}{I_\rho^{(\nu)}} = \left[\ln I_\rho^{(\mu)} - I\!\!E \ln I_\rho^{(\mu)}\right] - \left[\ln I_\rho^{(\nu)} - I\!\!E \ln I_\rho^{(\nu)}\right] \qquad (5.9)$$

Noticing that the function $\Phi_{N,\beta}[\omega](z)$ satisfies the Lipshitz bound

$$|\Phi_{N,\beta}[\omega](z) - \Phi_{N,\beta}[\omega'](z)| \leq \frac{1}{\sqrt{N}}\|\xi[\omega] - \xi[\omega']\|_2 \|z\|_2 \qquad (5.10)$$

we can again use Theorem 2.5, without this time, using its full power, given that the Lipshitz constant is bounded uniformly. This implies that for all $x \geq 0$

$$I\!\!P\left[\frac{1}{\beta N}\left(\ln I_\rho^{(\mu)} - I\!\!E \ln I_\rho^{(\mu)}\right) > x\right] \leq 4e^{-N\frac{x^2}{32(m^*)^2}} \qquad (5.11)$$

To complete the proof of Theorem 3 we show that with regard to the objects we consider, the measures $\widetilde{\mathcal{Q}}$ and $\mathcal{Q}$ differ only by exponentially small terms. More precisely

**Lemma 5.1:** *Assume that $\alpha \leq \gamma_a^2(m^*)^4$. Then on the set $\Omega_2$,*

$$\left|\mathcal{Q}_{N,\beta}(B_{c_5\gamma m^*}^{(\mu)}) - \widetilde{\mathcal{Q}}_{N,\beta}(B_{c_5\gamma m^*}^{(\mu)})\right| \leq e^{-c_8\beta M} \qquad (5.12)$$

**Proof:** From the fact that $\widetilde{\mathcal{Q}}$ is the convolution of $\mathcal{Q}$ with the Gaussian measure of mean zero and variance $\beta N$ it follows that

$$\mathcal{Q}(B_\rho^{(\mu)}) \leq \widetilde{\mathcal{Q}}(B_{\rho+\delta\gamma m^*}^{(\mu)}) + 2^M e^{-\frac{1}{4}\beta N\gamma^2\delta^2(m^*)^2} \qquad (5.13)$$

and

$$\mathcal{Q}(B_\rho^{(\mu)}) \geq \widetilde{\mathcal{Q}}(B_{\rho-\delta\gamma m^*}^{(\mu)}) - 2^M e^{-\frac{1}{4}\beta N\gamma^2\delta^2(m^*)^2} \qquad (5.14)$$

On the other hand,

$$\widetilde{\mathcal{Q}}(B^{(\mu)}_{\rho+\delta\gamma m^*}) - \widetilde{\mathcal{Q}}(B^{(\mu)}_{\rho-\delta\gamma m^*}) \leq \left[\frac{4\tilde{c}_-}{\tilde{c}_+}\right]^{M/2} e^{-\frac{\beta N \gamma^2 (m^*)^4 (c_5 - \delta)^2}{8\tilde{c}_+}} \tag{5.15}$$

by the same type of computation than the one leading to (5.7). Choosing $\delta = c_9 m^*$ with $c_9 > 1$, we obviously get (5.12) with $c_8$ depending only on $c_5$ and $\tilde{c}_+/\tilde{c}_-$. $\diamondsuit$

From Lemma 5.1 and (5.7) and (5.8) follows the first assertion of Theorem 3. The second follows from (5.9) and (5.11), provided

$$\sup_\mu \frac{e^{-\beta M c_8}}{\widetilde{\mathcal{Q}}(B^{(\mu)}_\rho)} \downarrow 0 \quad \text{a.s.} \tag{5.16}$$

as $N \uparrow \infty$. But clearly

$$\widetilde{\mathcal{Q}}(B^{(\mu)}_\rho) \geq \frac{1}{2M \inf_\nu I^{(\nu)}_\rho / I^{(\mu)}_\rho + J_\rho / I^{(\mu)}_\rho} \tag{5.17}$$

The second term in the denominator is exponentially small by (5.7) while by (5.11)

$$I\!P\left[2M \inf_\nu I^{(\nu)}_\rho / I^{(\mu)}_\rho \geq 2M e^{\beta M c_8 / 4}\right] \leq 4M e^{-\frac{\beta M (m^*)^2 c_8^2}{512}} \tag{5.18}$$

From here we get (5.16) and this concludes the proof of Theorem 3. $\diamondsuit\diamondsuit$

# Appendix: Proof of Lemma 4.18

We introduce the abbreviation $X_\mu \equiv X_\mu(N) = \frac{1}{\sqrt{N}} \sum_{j=1}^{N} \xi_j^\mu$. Lemma 4.18 can then be written in the following form

**Lemma A.1:** *Let $\delta_x$ denote the Dirac-measure concentrated on $x$ and let $\mathcal{N}_{0,\alpha}$ be the centered Gaussian measure with mean zero and variance $\alpha$. Then,*

$$w - \lim_{N \uparrow \infty} \frac{1}{N} \sum_{i=1}^{N} \delta_{(\xi_i, X)/\sqrt{N} - M/N} = \mathcal{N}_{0,\alpha} \quad I\!P\text{-a.s.} \tag{6.1}$$

**Proof:** To prove weak convergence, it is enough to prove the a.s. convergence on a measure determining class. The main step in the proof is thus the following lemma.

**Lemma A.2:** *Let $f \in C^{(2)}(I\!R)$ be an increasing bounded function with bounded first and second derivatives. Then*

$$\lim_{N \uparrow \infty} \frac{1}{N} \sum_{i=1}^{N} f\left(\frac{(\xi_i, X)}{\sqrt{N}}\right) - I\!E f\left(\frac{(\xi_i, X)}{\sqrt{N}}\right) = 0, \quad I\!P\text{-a.s.} \tag{6.2}$$

**Proof:** Use the exponential Chebeychev inequality to get that

$$
\begin{aligned}
&I\!P\left[\frac{1}{N}\sum_{i=1}^{N} f\left(\frac{(\xi_i,X)}{\sqrt{N}}\right) - I\!E f\left(\frac{(\xi_i,X)}{\sqrt{N}}\right) > \epsilon\right] \\
&\leq e^{-Nt\epsilon} I\!E \exp\left(t \sum_{i=1}^{N} f\left(\frac{(\xi_i,X)}{\sqrt{N}}\right) - I\!E f\left(\frac{(\xi_i,X)}{\sqrt{N}}\right)\right) \\
&= e^{-Nt\epsilon} \sum_x I\!E \left[\exp\left(t \sum_{i=1}^{N} f\left(\frac{(\xi_i,X)}{\sqrt{N}}\right) - I\!E f\left(\frac{(\xi_i,X)}{\sqrt{N}}\right)\right) \Big| X = x\right] I\!P[X = x] \\
&\equiv e^{-Nt\epsilon} \sum_x \widetilde{I\!E}_x \exp\left(t \sum_{i=1}^{N} f\left(\frac{(\xi_i,X)}{\sqrt{N}}\right) - I\!E f\left(\frac{(\xi_i,X)}{\sqrt{N}}\right)\right) I\!P[X = x]
\end{aligned}
\tag{6.3}
$$

where the last equality defines $\widetilde{I\!E}_x$. The crucial point is now that the variables $\xi_i$ under the law $\widetilde{I\!E}_x$ are negatively associated (see e.g.[JP,Lo]) and therefore

$$
\begin{aligned}
&\sum_x \widetilde{I\!E}_x \exp\left(t \sum_{i=1}^{N} f\left(\frac{(\xi_i,X)}{\sqrt{N}}\right) - I\!E f\left(\frac{(\xi_i,X)}{\sqrt{N}}\right)\right) I\!P[X = x] \\
&\leq \sum_x I\!P[X = x] \prod_{i=1}^{N} \widetilde{I\!E}_x \exp\left(tf\left(\frac{(\xi_i,X)}{\sqrt{N}}\right) - I\!E f\left(\frac{(\xi_i,X)}{\sqrt{N}}\right)\right) \\
&\leq \sum_x I\!P[X = x] \prod_{i=1}^{N} \left(1 + t[\widetilde{I\!E}_x f\left(\frac{(\xi_i,X)}{\sqrt{N}}\right) - I\!E f\left(\frac{(\xi_i,X)}{\sqrt{N}}\right)] + \frac{t^2}{2} \widetilde{I\!E}_x [f\left(\frac{(\xi_i,X)}{\sqrt{N}}\right) - I\!E f\left(\frac{(\xi_i,X)}{\sqrt{N}}\right)]^2 e^t\right)
\end{aligned}
\tag{6.4}
$$

where we have assumed, without loss of generality, that $|f(x)| \leq 1$. By the same hypothesis, the term proportional to $t^2$ in the last line is bounded by a constant, and we would immediately be done if the term proportional to $t$ was zero. While this is not exactly true, we will see that this is virtually true on a set of values $x$ which carry all but an exponentially small mass. Let us define

$$F(x) \equiv \widetilde{I\!E}_x f\left(\tfrac{1}{\sqrt{N}} \sum \xi_i^\mu x_\mu\right) \tag{6.5}$$

It is not difficult to show, using for instance the Yurinskii-martingale technique [Yu], that $F$ satisfies a concentration estimate.

**Lemma A.3:** *Le $F$ be defined by (6.5) and let $X = \tfrac{1}{\sqrt{N}} \sum_{i=1}^{N} \xi_i$. Then there exists a constant $0 < c_f < \infty$ such that for all $\delta < \alpha/2$,*

$$I\!P\left[|F(X) - I\!E F(X)| \geq \delta\right] \leq c_f \exp\left(-N \tfrac{\delta^2}{2c_f}\right) \tag{6.6}$$

**Proof:** Lemma A.3 is a concentration estimate for $F$ regarded as function of the $M$ independent random variables $X_\mu$. To get it, we will show that the derivative of $F$ with respect to $x_\mu$ satisfies appropriate bounds. We will use that the variables $\xi_i^\mu$ under $\widetilde{I\!E}_x$ are independent for different $\mu$ and that

$$\widetilde{I\!P}_x[\xi_i^\mu = \pm 1] = \tfrac{1}{2}\left(1 \pm \tfrac{x_\mu}{\sqrt{N}}\right) \tag{6.7}$$

Therefore, for any $\nu$ we can write

$$F(x) = \tfrac{1}{2}\widetilde{I\!E}_x \left[ f\left(\tfrac{1}{\sqrt{N}} \sum_{\mu \neq \nu} \xi_i^\mu x_\mu + \tfrac{x_\nu}{\sqrt{N}}\right) + f\left(\tfrac{1}{\sqrt{N}} \sum_{\mu \neq \nu} \xi_i^\mu x_\mu - \tfrac{x_\nu}{\sqrt{N}}\right) \right.$$
$$\left. + \tfrac{x_\nu}{\sqrt{N}} \left[ f\left(\tfrac{1}{\sqrt{N}} \sum_{\mu \neq \nu} \xi_i^\mu x_\mu + \tfrac{x_\nu}{\sqrt{N}}\right) - f\left(\tfrac{1}{\sqrt{N}} \sum_{\mu \neq \nu} \xi_i^\mu x_\mu - \tfrac{x_\nu}{\sqrt{N}}\right) \right] \right] \tag{6.8}$$

This representation allows immediately to compute the derivative with respect to $x_\nu$, and since $f$, $f'$ and $f''$ are assumed to be bounded, a simple computation shows that

$$\left|\tfrac{d}{dx_\nu} F(x)\right| \leq \tfrac{C|x_\nu|}{N} \tag{6.9}$$

This bound allows to estimate the conditional expectations

$$I\!E \left[ |X_\nu|^2 \left|\tfrac{d}{dX_\nu} F(X)\right|^2 e^{t|X_\nu|\left|\tfrac{d}{dX_\nu} F(X)\right|} \Big| \sigma(X_1, \ldots, X_{\nu-1}) \right] \leq \tfrac{C^2}{N^2} I\!E |X_\nu|^4 e^{t|X_\nu|^2/N} \tag{6.10}$$

where $\sigma(X_1, \ldots, X_{\nu-1})$ denotes the sigma algebra generated by the variables $X_1, \ldots, X_{\nu-1}$. Noting that $X_\mu$ are close to normal (and recalling e.g. Lemma 2.1), we see that the last expectation is bounded by $const./N^2$ as long as $2t/N < 1$ These allow the use of the Yurinskii-Martingale method (see e.g. [LT]; the specific computations used here will be similar to those in Chap 3 of [BGP2]) to prove (6.6). We leave the details to the reader. $\diamond$

As an immediate corollary of Lemma A.3 we get that except on a set of probability smaller that $c_f \exp(-\delta^2/2c_f)$,

$$|\widetilde{I\!E}_x f\left(\tfrac{(\xi_i,X)}{\sqrt{N}}\right) - I\!E f\left(\tfrac{(\xi_i,X)}{\sqrt{N}}\right)| \leq \delta \tag{6.11}$$

Therefore

$$\sum_x \widetilde{I\!E}_x \exp\left(t \sum_{i=1}^N f\left(\tfrac{(\xi_i,X)}{\sqrt{N}}\right) - I\!E f\left(\tfrac{(\xi_i,X)}{\sqrt{N}}\right)\right) I\!P[X=x]$$
$$\leq c_f \exp(-N\delta^2/2c_f) e^{tN} + \exp\left(+3Nt\delta + Nt^2 c_O\right) \tag{6.12}$$

for some finite constant $c_0$. Since for $t$ sufficiently small we can choose $\delta^2 = 2tc_f$, we may in fact use

$$\sum_x \widetilde{I\!E}_x \exp\left(t \sum_{i=1}^N f\left(\tfrac{(\xi_i,X)}{\sqrt{N}}\right) - I\!E f\left(\tfrac{(\xi_i,X)}{\sqrt{N}}\right)\right) I\!P[X=x] \leq c_f + \exp\left(N[3t^{3/2}\sqrt{c_f} + t^2 c_0/2]\right) \tag{6.13}$$

Inserting this bound in (6.3) and making, for $\epsilon$ sufficiently small, the choice

$$t = \frac{2}{81 c_f} \epsilon^2 \tag{6.14}$$

we get that for some finite constant $C$

$$I\!P\left[\tfrac{1}{N} \sum_{i=1}^N f\left(\tfrac{(\xi_i,X)}{\sqrt{N}}\right) - I\!E f\left(\tfrac{(\xi_i,X)}{\sqrt{N}}\right) > \epsilon\right] \leq \exp\left(-N \tfrac{2e^3}{81c_f}\right) \left[c_f + \exp\left(N \tfrac{2}{3} \tfrac{4e^3}{81c_f} + CN\epsilon^4\right)\right] \tag{6.15}$$

which for $\epsilon$ sufficiently small is of order $\exp(-Nc\epsilon^3)$. In much the same way we can also proof that

$$I\!P\left[\tfrac{1}{N}\sum_{i=1}^N f\left(\tfrac{(\xi_i,X)}{\sqrt{N}}\right) - I\!E f\left(\tfrac{(\xi_i,X)}{\sqrt{N}}\right) < \epsilon\right] \leq \exp(-Nc\epsilon^3) \tag{6.16}$$

Form this the lemma follows by the Borel-Cantelli Lemma.$\diamond$

To conclude we have to identify $\lim_{N\uparrow\infty} I\!E f\left(\tfrac{(\xi_i,X)}{\sqrt{N}}\right)$. Clearly this quantities are the same for all $i$ and

$$\tfrac{(\xi_1,X)}{\sqrt{N}} = \tfrac{1}{N} \sum_\mu \sum_{j\neq 1} \xi_1^\mu \xi_j^\mu + \tfrac{M}{N} \tag{6.17}$$

the central limit theorem applied to the independent random variables $\{\xi_1^\mu \xi_j^\mu\}_{j\geq 2, \mu \geq 1}$ shows that

$$\lim_{N\uparrow\infty} I\!E f\left(\tfrac{(\xi_i,X)}{\sqrt{N}} - \tfrac{M}{N}\right) = \tfrac{1}{\sqrt{2\pi\alpha}} \int dz f(z) e^{-\tfrac{z^2}{2\alpha}} \tag{6.18}$$

This together with Lemma A.2 implies Lemma A.1. $\diamond$